\documentstyle[11pt,aaspp4,tighten]{article}
\def\gtaprx {\lower .1ex\hbox{\rlap{\raise .6ex\hbox{\hskip .3ex
             {\ifmmode{\scriptscriptstyle >}\else
                {$\scriptscriptstyle >$}\fi}}}
                \kern -.4ex{\ifmmode{\scriptscriptstyle \sim}\else
                {$\scriptscriptstyle\sim$}\fi}}}
\def\ltaprx {\lower .1ex\hbox{\rlap{\raise .6ex\hbox{\hskip .3ex
             {\ifmmode{\scriptscriptstyle <}\else
                {$\scriptscriptstyle <$}\fi}}}
                \kern -.4ex{\ifmmode{\scriptscriptstyle \sim}\else
                {$\scriptscriptstyle\sim$}\fi}}}
\def\etal {et al. }
\def\littleprime{\ifmmode{\scriptscriptstyle \prime }
    \else{\hbox{$\scriptscriptstyle \prime$ }}\fi}
\def\littless{\ifmmode{\scriptscriptstyle s }
    \else{\hbox{$\scriptscriptstyle s $ }}\fi}
\def\littlemm{\ifmmode{\scriptscriptstyle m }
    \else{\hbox{$\scriptscriptstyle m $ }}\fi}
\def\littlehh{\ifmmode{\scriptscriptstyle h }
    \else{\hbox{$\scriptscriptstyle h $ }}\fi}
\def\littlecirc{\ifmmode{\scriptscriptstyle \circ }
    \else{\hbox{$\scriptscriptstyle \circ $ }}\fi}
\def\rasec{\raise .9ex \hbox{\littless}}
\def\arcsec{\raise .9ex
            \hbox{\littleprime\hskip-3pt\littleprime\hskip-3pt}}
\def\ramin{\raise .9ex \hbox{\littlemm}}
\def\arcmin{\raise .9ex \hbox{\littleprime}}
\def\hrs{\raise .9ex \hbox{\littlehh}}

\def\degree{\raise .9ex \hbox{\littlecirc}}
\def\magpoint{\hbox to 2pt{}\rlap{\hskip -.5ex \arcmm}.\hbox to 2pt{}}
\def\arcsspoint{\hbox to 1pt{}\rlap{\arcss}.\hbox to 2pt{}}
\def\arcsecpoint{\hbox to 1pt{}\rlap{\arcsec}.\hbox to 2pt{}}
\def\arcminpoint{\hbox to 1pt{}\rlap{\arcmin}.\hbox to 2pt{}}
\def\degreepoint{\hbox to 1pt{}\rlap{\degree}.\hbox to 2pt{}}

\slugcomment{Accepted by {\it The Astrophysical Journal Supplements}.}

\begin{document}

\title{The Nature of Composite LINER/H\,II Galaxies, As Revealed from 
High-Resolution VLA Observations}

\author{Mercedes E. Filho}
\affil{Lisbon Observatory and Kapteyn Astronomical Institute}
\authoraddr{P.O.~Box 800, NL--9700 AV Groningen, The~Netherlands}

\author{Peter D. Barthel}
\affil{Kapteyn Astronomical Institute}
\authoraddr{P.O.~Box 800, NL--9700 AV Groningen, The~Netherlands}

\author{Luis C. Ho}
\affil{Observatories of the Carnegie Institution of Washington}
\authoraddr{813 Santa Barbara Street, Pasadena, CA 91101, USA}

\begin{abstract}

A sample of 37 nearby galaxies displaying composite LINER/H~II and pure
H~II spectra was observed with the VLA in an investigation of the nature
of their weak radio emission.  The resulting radio contour maps overlaid
on optical galaxy images are presented here, together with an
extensive literature list and discussion of the individual galaxies.
Radio morphological data permit assessment of the ``classical AGN''
contribution to the global activity observed in these ``transition'' LINER
galaxies. One in five of the latter objects display clear AGN
characteristics: these occur exclusively in bulge-dominated hosts.

\end{abstract}


\section{Introduction}

LINER galaxies are a class of active galaxies characterized by the presence of 
strong, nuclear, low-ionization emission lines (Heckman 1980).  Like 
Seyfert nuclei, LINERs emit much stronger optical low-ionization forbidden 
lines compared to H~II nuclei, whose line emission is powered by 
photoionization from young massive stars, but LINERs have a characteristically
lower ionization state than Seyferts.  In a recent optical survey of nearly 
500 nearby galaxies, Ho, Filippenko, \& Sargent (1995, 1997a, 1997b) find that
$\sim$20\% of all galaxies brighter than $B_T$ = 12.5\,mag display
LINER-type spectra.  An additional 13\% of the objects show spectra 
intermediate between those of ``pure'' LINERs and ``pure'' H\,II nuclei. 
Ho, Filippenko, \& Sargent (1993; see also Ho 1996) dubbed this class as 
``transition objects,'' and they hypothesized that these might be composite 
systems in which the optical signal of a weak active nucleus (the LINER 
component) has been spatially blended by circumnuclear star-forming regions 
(the H~II component).

Models have shown that photoionization by hot stars are able to reproduce the
optical spectra of LINERs (Terlevich \& Melnick 1985), especially those of 
the transition-type variety (Shields 1992; Filippenko \& Terlevich 1992).  
Other researchers have advocated shock-wave heating as a viable excitation 
mechanism for LINERs (Fosbury \etal 1978; Heckman 1980; Dopita \& Sutherland 
1995).  At the same time, an increasing body of evidence suggests that a 
significant fraction of LINERs do contain irrefutable AGN characteristics 
(see reviews by Filippenko 1996 and Ho 1999b).  Radio surveys, in particular, 
have shown that many LINERs exhibit weak nuclear radio emission (Heckman 1980;
Sadler, Jenkins, \& Kotanyi 1989; Wrobel \& Heeschen 1991; Slee \etal 1994; 
Falcke, Wilson, \& Ho 1997; Van Dyk \& Ho 1998).  As discussed in detail by
Ho (1999a), weak nuclear radio emission in early-type (elliptical and S0) 
galaxies is most likely related to accretion-driven energy production in these 
objects, but at a very low level compared to powerful radio galaxies, which 
are also usually hosted by early-type systems.  

Radio observations provide an expecially attractive tool to further our 
understanding of the LINER phenomenon.  Since radio frequencies are not plagued
by dust obscuration and photoelectric absorption, high-resolution radio
observations potentially offer the cleanest and most efficient method to
isolate an AGN core.  Although radio emission also arises from supernova
remnants and H\,II regions, AGN cores can be identified through their 
high degree of central concentration, small angular size, flat spectral 
indices, and high brightness temperatures.

Given that LINERs are so numerous, they may be the most common type of
AGN known.  Their large space densities could make a major impact on the faint 
end of the local AGN luminosity function, which in turn has ramifications for 
many astrophysical issues ranging from the cosmological evolution of AGN to
the contribution of AGN to the cosmic X-ray background.  We are
therefore conducting high-resolution radio observations of selected
LINER samples in order to investigate the interrelation between LINERs
and classical AGN.  For the present paper --- the first in a series ---
we have concentrated on transition-type LINERs.  We report on sensitive 
Very Large Array (VLA)\footnote{The VLA is a facility of the National Radio 
Astronomy Observatory (NRAO) which is a facility of the National Science 
Foundation operated under cooperative agreement by Associated Universities, 
Inc.} C-array observations of a sample of 25 such transition LINERs at 
8.4\,GHz.  The resulting images, having angular resolution of about 
2\arcsecpoint5, allow us to assess the strength and spatial distribution of 
the nuclear radio emission.  For comparison, we have also observed a small
sample of 12 pure H\,II nuclei. 

In the following discussion and in upcoming papers, we adopt a
Hubble constant of H$_{0}$ = 75~km~s$^{-1}$~Mpc$^{-1}$.

\section{Sample Selection}

The sample of 25 transition LINER/H\,II galaxies and 12 pure H\,II
nuclei was extracted from the original magnitude-limited Palomar survey of 
486 bright northern galaxies (Ho \etal 1995), following the classification 
criteria outlined in Ho et al. (1997a).  The Palomar sample contains 65 
transition nuclei (Ho et al. 1997a), from which we chose a subset of 25 that 
had little or no arcsec-scale resolution radio continuum data published, and
that fell within the observation window assigned to us by the VLA.  In 
addition, we selected for comparison a small sample of 12 pure H\,II nuclei 
(out of 206 such objects in the full Palomar sample).  These were chosen to
mimic the Hubble type distribution of the sample of transition objects, 
again taking into consideration scheduling constraints.
We are not aware of any strong biases introduced by our sample selection. 

Table~1 lists the 37 sample objects.  In addition to the optical positions of 
the galaxies (columns 3 and 4), this table lists their distance (column~5) 
and their Hubble type (column~6).  The Hubble types and distances are taken 
from Ho \etal (1997a).  The ``reference'' column (7), together with Table~2,
supplies references to earlier radio studies; it is readily apparent that
several of the sample objects are well-studied objects, while some are
not.  We will discuss our observational results in light of these existing 
data in Section~4.  The sample galaxies span a wide range of Hubble types, but 
the majority are spirals.

\medskip \centerline{\bf -- TABLEs~1/2: Sample of transition LINER
  galaxies and H\,II nuclei --}

\section {Observations and Data Reduction}

The resulting sample of 25 composite LINER/H\,II and 12 pure H\,II galaxies was
observed with the X-band (8.4\,GHz) system of the VLA in two
observing sessions, on 1997 August~21 and August~25.  The array was in its
C-configuration, yielding a typical resolution of 2\arcsecpoint5. Two
bands of 50\,MHz each were combined, yielding a total bandwidth of
100\,MHz. Snapshot observations combining two scans of 8--9~minutes
each, at different hour angles, were obtained in order to improve the
shape of the synthesized beam.

Secondary phase and amplitude calibrators were observed before and after
each scan.  The primary calibrator was 3C\,286, with appropriate
baseline constraints, and the adopted 8.4\,GHz flux densities, as provided
by the VLA staff, were 5.22~Jy and 5.20~Jy, at IF1 and IF2, respectively. 
The calibration uncertainty is dominated by the uncertainty in the
absolute flux density of the primary calibrator, which is a few percent.  
The array performed well: judging from the calibration sources,
the antenna phase and amplitude calibration appeared stable to within a
few percent. 

The radio data are of high quality, and there was no need for
extensive flagging of discrepant points. Reduction of the data was
performed using standard NRAO AIPS (version 15OCT98) image processing
routines.  AIPS task IMAGR was employed to Fourier transform the
measured visibilities and obtain CLEAN (H\"ogbom 1974; Clark 1980) maps of the sources. 
Full-resolution
maps, having synthesized beams of roughly 2\arcsecpoint5, and tapered
maps, having beams between about 5\arcsec~and 12\arcsec, were
made. This proved useful to detect weak low surface brightness
emission. As a rule of thumb, 1\arcsec~corresponds to 50--100~pc for the 
typical distances involved. Most images reach
the theoretical noise level of $\sim 0.04$~mJy/beam (Perley, Schwab, \& Bridle 
1989) to within a factor of two.  Self-calibration was employed in
the analysis of two of the stronger sources (NGC\,4552 and NGC\,5354), 
leading to considerable improvements.

\section {Results}

\subsection {Radio Maps}

Noting that a non-detection should be taken to imply no correlated flux
density in excess of 0.5~mJy on the arcsec scale, radio contour maps of
the detected galaxies are presented in Figures 1--7, at the end of the
paper.  The radio maps are overlayed on optical images taken from the
Digitized Sky Survey (DSS).  In a number of cases only the
high-resolution map (typical beamsize 2\arcsecpoint5) or the
low-resolution map (typical beamsize 7\arcsec~-- 10\arcsec) are shown. 
Unrelated background radio sources were found in several fields; their
number is entirely as expected given the size of the primary beam and
the sensitivity of the observations.  In cases of large distances from
the field center, primary beam corrections were applied.  Some of the
background sources have been identified using NED (NASA/IPAC
Extragalactic Database) and the Cambridge APM (Automatic Plate
Measuring) facility.  Where possible, we have included them by their
name in the tables below.  Previously uncatalogued background sources
are designated by their sky orientation with respect to the field center
for the relevant galaxy.  All maps use contouring according to CLEV
$\times $ (--3, 3, 6, ..., MLEV) mJy/beam where MLEV are powers $3
\times 2^{\rm n}$ up to a maximum of 96 and CLEV is the typical noise
level on each radio map (see Table~3). 

Table~3 lists the various map parameters, with the following column
headings: the source name (1), the applied taper (2), the resulting
restoring beam (3), the position angle of the beam (4), the image noise 
level (5), and the relevant figure number (6). The H\,II nuclei 
appear at the bottom of the table.

\medskip \centerline{\bf -- TABLE~3: Radio map parameters --}

Given that good phase stability was obtained, as judged from the VLA phase
calibrators, the astrometric accuracy of our overlay procedure must be
dominated by the DSS accuracy, which is known to be about 0.6\arcsec~ 
(V\'eron-Cetty \& V\'eron 1996). This is born out by the images of several
unresolved and slightly resolved nuclei (NGC\,4552, NGC\,5354, 
NGC\,5838, NGC\,5846) which appear accurate to within a few tenths of an arcsec.  

\subsection {Comments on Individual Galaxies}

With reference to the images in the Appendix, we proceed by discussing
the results of our VLA imaging in the light of earlier studies of the
sample objects.  This discussion also deals with the detected background
sources and includes important bibliographic information.  We will
frequently compare our measurements with sample galaxy flux densities
obtained within the VLA 1.4\,GHz NVSS and FIRST surveys, having typical
resolution of 45\arcsec~ and 5\arcsec, respectively (Condon \etal 1998a;
Becker, White, \& Helfand 1995), as well as with the Green Bank 20~cm
and 6~cm surveys, having typical resolution of $\sim$700\arcsec~and
3\arcminpoint5, respectively (White \& Becker 1992, WB92 hereafter;
Becker, White, \& Edwards 1991, BWE91 hereafter). 

{\bf IC\,520 :} No compact structure was seen in this object, not even
on the tapered map.  It was also not detected in the NVSS.  However,
Condon (1987) reports a detection at 1.49\,GHz: at
0\arcminpoint8~resolution, he measured a 3.1~mJy radio source. 

{\bf NGC\,2541 :} We did not detect compact structure, not even on the
tapered map. This source was also not detected in the FIRST or in the
NVSS survey. However, Condon (1987) reports a possible detection at 
1.49\,GHz, 0\arcminpoint8~resolution, of a 3.2~mJy source, slightly
extended to the SW relative to the optical position of the galaxy.

{\bf NGC\,2985 :} We detect a 1.9~mJy source.  At
2\arcsecpoint5~resolution, the source is slightly extended in the N-S
direction, along the orientation of the optical galaxy.  The tapered map
shows a 2.7~mJy core and about 1~mJy weak low surface brightness
emission to the SW.  The object was also detected in the NVSS: 44.1~mJy. 
At 1.49\,GHz, 1\arcminpoint0~resolution, Condon (1987) measured a total
flux of 61.9~mJy: the radio emission of NGC\,2985 must be extended over
tens of arcseconds. 

{\bf NGC\,3593 :} On both the untapered and tapered maps we measure
about 20~mJy of diffuse, 45\arcsec~E-W elongated emission, oriented
along the major axis of the host of this H\,II nucleus.  NVSS measured
87~mJy emission, WB92 measured 132~mJy (hence the radio source exceeds
one arcminute in size), whereas BWE91 measured 53~mJy. The 1.4\,GHz,
5\arcsec~resolution map of Condon \etal (1990) shows an E-W oriented
source of 63.4~mJy: these and our VLA data imply a steep-spectrum
radio source.

We also detect an unrelated 10.6~mJy source, 3\arcmin~to the N, extended
towards the E. As judged from the APM survey, the POSS plates show an as 
yet unclassified nonstellar source at this position: background source
NGC\,3593N is most likely a weak distant radio galaxy.

{\bf NGC\,3627, M\,66 :} This interacting galaxy belongs to the Leo
triplet.  The 8.4\,GHz maps show a 2\arcmin~triple source aligned with
the inner bar of the optical galaxy. The outer radio components are
related to star formation in the disk of the galaxy (Urbanik, Gr\"ave, 
\& Klein 1985). We measure 3.9~mJy integrated emission in the compact central
component, while $\sim$25~mJy are distributed in the NW and SE components.
From comparison with low-resolution surveys (NVSS: 324.9~mJy; WB92:
434~mJy; BWE91: 141~mJy) we infer that most of the radio emission of
this steep-spectrum radio source is resolved out by our observations.
The triple radio structure was also observed by Saikia \etal (1994),
at 5\,GHz with 2\arcsec~resolution. Combining our data with the Saikia
\etal (1994) and the Hummel \etal (1987) data, we conclude that the
nuclear radio source in NGC3627 must be of variable, flat-spectrum
nature. More recently, high resolution (0\arcsecpoint15) 15\,GHz
observations detected an unresolved 1.4~mJy core (N. Nagar, private
communication).

{\bf NGC\,3628 :} Like NGC\,3627, this galaxy also belongs to the Leo
triplet.  Our 8.4\,GHz full-resolution map shows some extended emission
to the N and an eastern extension that is roughly aligned with the
projected disk of the galaxy.  The nuclear region is clearly resolved. 
The total flux density on our low-resolution map is about 70~mJy, 70$\%$
of which is in $\sim$15\arcsec~diffuse emission.  The NVSS measured
292~mJy whereas WB92, at lower resolution, detected 402~mJy, indicating
resolved large-scale emission.  At 1.4\,GHz Condon \etal (1990) measured
203~mJy and 205~mJy at 5\arcsec~and 1\arcsecpoint5~resolution,
respectively.  These and our data imply a steep-spectrum radio source,
in agreement with the results from the Effelsberg 100m telescope
(Schlickeiser, Werner, \& Wielebinski 1984).  Carral, Turner, \& Ho
(1990) observed this object at 15\,GHz using the VLA in A-array and
measured a total flux of 23~mJy.  These authors sampled the innermost
part of the source and obtain a $\sim4$\arcsec~string of a dozen
components aligned with the major axis of the galaxy, that they suggest
could be star-forming regions in the disk. 

{\bf NGC\,3675 :} We detect about 1.5~mJy of low surface brightness
emission on our tapered maps.  NVSS detected a 48.9~mJy source, whereas
FIRST measured 8.04~mJy: the source must be strongly resolved. 
1.49\,GHz VLA D-array observations (Condon 1987; Gioia \& Fabbiano 1987)
detected a $\sim$50~mJy source, oriented N-S, along the major axis of
the optical galaxy.  Condon, Frayer, \& Broderick (1991) report a flux
density of 27~mJy at 4.85\,GHz and a steep spectral index between 4.8
and 1.4\,GHz. 

{\bf NGC\,3681 :} We did not detect a radio source on the untapered or
tapered maps. In contrast, NVSS measured a weak source of about
4.2~mJy -- at much lower resolution.

{\bf NGC\,3684 :} This H\,II nucleus had no radio source detected,
either on the untapered or on the tapered maps, but NVSS detected a
15.9~mJy source.  The radio emission of NGC\,3684 must be diffuse.

We do, however, detect a partially resolved 18.1~mJy background
source, 3\arcmin~to the SW, which APM identifies with a 19.5 mag
stellar-like object. Further identification is still lacking.

{\bf NGC\,4013 :} We detect a moderately resolved 1.1~mJy core and a small
extension to the NE, parallel to the projected disk of this edge-on
galaxy.  There is an additional 3~mJy of weak disk emission.  On the
basis of the widely different NVSS and FIRST detections (40.5 vs. 
11.9~mJy), this disk emission must extend tens of arcseconds.  Hummel,
Beck, \& Dettmar (1991) present 10\arcsec~resolution 5\,GHz data of
NGC\,4013 (UGC\,6963) which show both the prominent core and the
extended disk emission. Recent high resolution (0\arcsecpoint15) 15\,GHz
observations did not detect NGC\,4013 above a 10$\sigma$ limit of 1~mJy
(N.~Nagar, private communication).

We also detect a slightly resolved 3.8~mJy background source,
2\arcmin~to the far SE.  This source is also clearly seen in the map of
Condon (1987).  The APM facility reveals a 19.6 mag stellar-like object,
which lacks further identification as yet. 

{\bf NGC\,4100 :} We detect about 7~mJy of low surface brightness
emission associated with this H\,II nucleus.  The tapered map shows a
slight extension to the SE along the major axis of the optical
galaxy. NVSS detected a 50.3~mJy source and FIRST detected 17.8~mJy,
indicating resolution effects.

{\bf NGC\,4217 :} We detect about 22~mJy, distributed along
$\sim$1\arcminpoint5~of the projected galactic disk of this H\,II
nucleus. From low-resolution observations (NVSS: 123~mJy; WB92:
139~mJy; BWE91: 40~mJy) we conclude that NGC\,4217 harbours extended
low surface brightness radio emission, which must have a steep radio
spectrum.
 
{\bf NGC\,4245 :} No radio source associated with this H\,II nucleus
was detected on the untapered or tapered maps. Neither NVSS nor FIRST
has detected this object at 1.4\,GHz.

{\bf NGC\,4321, M\,100 :} We detect about 16~mJy of low surface
brightness emission in the nuclear region of this well-known grand
design spiral in the Virgo cluster.  The tapered map shows emission
slightly extended to the NW. NVSS detected 87~mJy, while WB92 measured
323~mJy and BWE91 87~mJy.  At 1.49\,GHz, 0\arcminpoint9~resolution
Condon (1987) measured 180\,mJy total flux in a $\sim $ 3\arcmin
$\times $ 2\arcmin region.  At 4.9\,GHz, 1\arcsecpoint5~resolution
Collison \etal (1994) detect a ring-like structure coincident with our
main feature.  At 8.5\,GHz, 0\arcsecpoint2 resolution, these authors
detect two unresolved sources, having peaks of 0.37~mJy/beam (E) and
0.22\,mJy/beam (W), respectively. The eastern radio source is
coincident with their 4.9\,GHz main component while the one to the W
is coincident with the optical nucleus and has a flat spectrum
(Collison \etal 1994).

We also detect an unresolved source, 1.5\arcmin~to the SE, with 1.3~mJy
integrated flux, apparently located near the middle of the southern
spiral arm of the galaxy.  It is clearly visible on the NGC\,4321 map. 
We have identified this object with SN\,1979C (Weiler \etal 1981). 
Collison \etal (1994) measured 2~mJy for this SN\,1979C, at 4.9\,GHz. 

{\bf NGC\,4369, Mrk\,439 :} We detect 3.8~mJy of low surface brightness
disk emission associated with this H\,II nucleus.  On the basis of
widely different NVSS and FIRST detections (24.3 vs.  4.67\,~mJy), this
disk emission must extend over tens of arcsec.  The Condon \etal (1990)
1.49\,GHz image, at 15\arcsec~resolution, shows a 18.3~mJy source with
extended emission to the E. 

{\bf NGC\,4405, IC\,788 :} No compact emission was detected in the
untapered or tapered maps.  NVSS detected weak (4.5~mJy) emission for
this H\,II galaxy. 

We did, however, detect a partially resolved 10.1~mJy source,
2\arcmin~SW of the NGC\,4405 target position.  As judged from the APM,
the POSS plates show a 19.2 mag stellar-like object at this position,
which we subsequently identify as the $z=1.929$ QSO, LBQS\,1223+1626
(e.g., Hewett, Foltz, \& Chaffee 1995). 

{\bf NGC\,4414 :} We detect $\sim$29~mJy of low surface brightness
emission, distributed along $\sim$1\arcminpoint5, in a structure aligned
with the galaxy's major axis.  NVSS detected a 242~mJy source while
FIRST detected a double-peaked source with 44 and 64~mJy components. 
The Condon (1983), 1.465\,GHz, 12\arcsecpoint5~resolution map and the
Condon (1987), 1.49\,GHz, 1\arcminpoint0~resolution map show comparable
N-S structure, measuring $\sim$0.22~Jy.  Using this last value plus the
78~mJy measured at 4.85\,GHz, 15\arcsec~resolution, Condon \etal (1991)
obtain a steep spectral index. 

{\bf NGC\,4424 :} On the tapered map we detect a resolved $\sim$1~mJy core 
plus 1.5~mJy weak low surface brightness emission to the E along the
projected disk of this candidate merger galaxy (Kenney \etal 1996). 
NVSS detected a weak 4.5~mJy source.  Our full-resolution data show the
core of this H\,II nucleus to be somewhat resolved. This is borne out of
our recent high resolution
(0\arcsecpoint25) 8.4\,GHz observations which did not detect NGC\,4424 above
a 3$\sigma$ limit of 0.2~mJy.

We also detect an unrelated, slightly resolved 13.3~mJy source,
2\arcminpoint5~to the SE.  The APM identifies it with a 20.2 mag
stellar-like object.  Further properties of this background source
remain as yet unknown. 

{\bf NGC\,4470 :} We detect $\sim$3~mJy of weak low surface
brightness emission, extending 30\arcsec~along the galaxy major 
axis. NVSS detected 17.1 mJy associated with this H\,II nucleus.

We also detect an unrelated, resolved 16.0~mJy radio source,
1\arcminpoint5~to the NE of NGC\,4470, which is not identified on the
POSS plate (APM).  We identify this object with the radio source
TXS\,1227+081 (Douglas \etal 1996), which has as yet no optical
identification.

{\bf NGC\,4552, M\,89 :} We detect about 77~mJy in a strong core plus 
possibly a jet-like structure to the NE.  NVSS
detected a 103~mJy source.  Since BWE91 measured 64~mJy at 5\,GHz, the
radio source in NGC\,4552 must have a relatively flat radio spectrum,
which is in agreement with findings by Condon \etal (1991).  The radio
source appears to be variable (Wrobel \& Heeschen 1984; Ekers, Fanti, \&
Miley 1983; Sramek 1975a, 1975b; Ekers \& Ekers 1973). Our recent
subarcsec resolution 8.4\,GHz images have confirmed this object's compactness.  

{\bf NGC\,4643 :} No compact radio structure was detected on the
untapered or the tapered maps.  NVSS also did not detect this source. 
However, we have detected several very weak sources ($\sim$1~mJy) on the
tapered maps, some of which may be due to weak diffuse disk emission. 

{\bf NGC\,4710 :} We detect about 6~mJy of weak disk emission
associated with this H\,II nucleus.  NVSS measured a 19.3~mJy source.
At 1.49\,GHz, Condon \etal (1990) measure 17.2~mJy and 14.7~mJy at 15
and 5\arcsec~resolution, respectively. The data imply a resolved
steep-spectrum radio source.

We also detect a 4.8~mJy resolved background source, 2.5\arcmin~to the
E of NGC\,4710. No indentification could be found on the POSS.

{\bf NGC\,4713 :} We detect $\sim$1.2~mJy weak low surface brightness disk
emission on our tapered maps. NVSS measured a 46.9~mJy source: our 
observations must have resolved out much of the flux.

{\bf NGC\,4800 :} We detect about 1.2~mJy of weak low surface brightness
emission, stretching $\sim30$\arcsec~along the galaxy major axis.  NVSS
detected a 23.5~mJy source and FIRST 12.2~mJy.  Therefore, our
observations are resolving out much of the emission of this H\,II
nucleus. 

{\bf NGC\,4826 , M\,64:} We detect about 21~mJy of low surface
brightness emission.  The central region on the full-resolution map (not
presented here) shows a double-peaked component, which can be compared
to the complex inner triple structure of the 15 and 5\,GHz,
2\arcsec~resolution maps of Turner \& Ho (1994).  Combining the data
indicates that this is a steep spectrum source.  NVSS detected a 101~mJy
source and BWE91 detected a 56~mJy source.  The 1.49\,GHz,
1\arcminpoint0~resolution data of Condon \etal (1998b) and Condon (1987)
show a $\sim$100~mJy source, as do the data of Gioia \& Fabbiano (1987)
at 40\arcsec~resolution. 

{\bf NGC\,4845 :} On our tapered maps we detect a 2~mJy core and some
emission to the N, perpendicular to the galaxy major axis, plus 10~mJy
of $\sim20$\arcsec~elongated disk emission.  NVSS detected a 43.8~mJy
source associated with this H\,II nucleus.  The data of Condon \etal
(1990) imply resolution effects on the 10~arcsec scale. 

{\bf NGC\,5012 :} We detect some very weak ($\sim$1~mJy) features
near the target phase center. NVSS detected a 31.4~mJy source, but 
as FIRST did not detect it, we must be dealing with extended low
surface brightness emission.

{\bf NGC\,5354 :} We detect an unresolved 11.7~mJy core.  NVSS and
FIRST detected a 8.4 and 8.0~mJy source, respectively, implying that the
NGC\,5354 radio source is unresolved at a resolution of 5\arcsec.  The
4.85\,GHz, 15\arcsec~resolution maps of Condon \etal (1991) show 7~mJy
emission, implying that the nuclear radio source has an inverted
spectrum.  Our recent
subarcsec resolution 8.4\,GHz images have confirmed this object's compactness.

Paired galaxy NGC\,5353 is also detected, 1\arcminpoint25~to the S,
with a 26.7~mJy core. Condon \etal (1991) measure 29~mJy at 4.85\,GHz,
15\arcsec~resolution, implying that NGC\,5353 also hosts a flat
spectrum radio source.

{\bf NGC\,5656 :} We detect $\sim$1~mJy of weak low surface brightness
emission, stretching over $\sim$25\arcsec~along the major axis of the
galaxy. NVSS detected a 22~mJy source.

{\bf NGC\,5678 :} We detect about 8.5~mJy of low surface brightness
emission stretching over $\sim$1\arcminpoint2~along the major axis of
the galaxy.  The tapered map shows what could be a double-peaked source,
similar to the 6\arcsec~resolution, 1.4\,GHz map of Condon (1983).  NVSS
detected a 112~mJy source while BWE91 measured 68~mJy.  The NVSS and
BWE91 results, combined with Condon (1983) reporting 109~mJy at
1.4\,GHz, imply a steep spectrum source. 

{\bf NGC\,5838 :} We have detected a slightly resolved 2.2~mJy source. 
NVSS detected 3.0~mJy, while Wrobel \& Heeschen (1991) report a 2 mJy
source (5\,GHz, 5\arcsec~resolution): the nuclear radio source in
NGC\,5838 must have a flat radio spectrum.  Our recent
subarcsec resolution 8.4\,GHz images have confirmed this object's compactness.

There is also an unrelated, unresolved 1.3~mJy source, about
1\arcminpoint5~to the S of NGC\,5838.  The POSS plates reveal an 18.4
mag star-like object at that position, for which as yet no redshift is
available. 

{\bf NGC\,5846 :} We have detected a partially resolved 7.1~mJy source. 
NVSS measured a 22~mJy source.  1.4\,GHz VLA observations by
M\"ollenhoff, Hummel, \& Bender (1992) measured an unresolved core of
9~mJy plus 10~mJy of additional diffuse emission: NGC\,5846 must possess
a compact flat-spectrum core component. Our recent
subarcsec resolution 8.4\,GHz images have confirmed this object's compactness.

{\bf NGC\,5879 :} No radio structure was detected on the untapered or
the tapered maps. NVSS detected a 21~mJy source, which hence must arise 
in an extended low surface brightness region.

However, we did detect a 292.1~mJy, slightly resolved background
source, about 2\arcmin~NE of our target. It shows up only on the POSS
red plate as a noise-like source. We have identified this source as
QSO 1508+5714, at a redshift of 4.301 (Hook \etal 1995). It is in the
WENSS (Rengelink \etal 1997) and in BWE91 (282~mJy) as well as in WB92
(149~mJy). Patnaik \etal (1992) have observed the source with the VLA
A-array at 8.5\,GHz and measured 153~mJy: 1508+5714 must be a variable 
flat-spectrum quasar.
 
{\bf NGC\,5921 :} We have detected a very weak 0.5~mJy core and some
additional weak extended emission.  At 1.49\,GHz and 0\arcminpoint9
resolution, Condon (1987) measured a 20.8~mJy source with an
additional 2.8~mJy eastern component which we do not detect at
8.4\,GHz. All emission must be of low surface brightness nature.

{\bf NGC\,6384 :} Following the NVSS non-detection, we also do not
detect (compact) radio emission from NGC\,6384.  However, Condon
(1987), at 1.49\,GHz and 1\arcminpoint2~resolution, found the radio
source, having $\sim$ 35~mJy, to be very extended.

{\bf NGC\,6482 :} We have not detected this source at 8.4\,GHz and
neither did NVSS.

We did, however, detect a slightly resolved 23.6~mJy background
source, about 3\arcmin~to the S of NGC\,6482. On the full-resolution
map this source is slightly extended. The POSS plates show a 20.9 mag
stellar object. We have identified this source with the radio source
1749+2302 included in GB6, with 26~mJy flux density.

{\bf NGC\,6503 :} We do not detect radio emission from this galaxy on
the full-resolution or on the tapered maps.  1.4\,GHz data (NVSS and Condon
1987) indicate a NW-SE extended $\sim$40~mJy source, which
consequently must be diffuse.
 
We also detect a background source, about 6\arcmin~to the SW.  Due to
the large distance from the phase center, which implies a large and uncertain
primary beam correction, we cannot assess the flux density of this
source.  The source appears as a 16.9 mag stellar-like object on the POSS
plates.  We have identified this object with the BL Lacertae source
1749+701 at redshift 0.77 (Hughes, Aller, \& Aller 1992).  This object has been
extensively observed at other radio-frequencies and is included in the
WENSS (Rengelink \etal 1997) as well as BWE91 and WB92.  The only other
8.5\,GHz measurement of this source comes from Patnaik \etal (1992).  They
measure 558~mJy with the VLA A-array. 

Table~4 summarizes the properties of the background sources, including
the previously identified, as well as the new, identifications. Tabulated 
8.4\,GHz fluxes have been corrected for primary beam attenuation. 

\medskip \centerline{\bf -- TABLE~4: Field/background sources --}

\subsection {Radio Source Parameters}

For each galaxy with a 8.4\,GHz detection, we list in Table~5 the radio
source parameters, measured both from the full-resolution and the
tapered maps. Peak and integrated values (columns~3 and 6), as well as 
well as sky positions of bidimensional gaussian fits (columns~4 and 5)
are tabulated. These values were obtained using the AIPS task IMFIT.  
As we are often dealing with resolved and/or asymmetric emission, such 
measurements are often inaccurate and must represent lower limits.  
In such cases, we have estimated the integrated
8.4\,GHz flux density from inspection of the map and the relevant clean
component file and we have given it the prefix ``$\geq$'' (column~6). 
Also, given the fact that we have fitted single gaussian to the brightest
source components, the accuracy of the radio peak positions is judged to be
\ltaprx1\arcsec~(see images). For NGC\,3627, the quoted peak and integrated 
value are of the central core.  In parentheses we include the total flux 
of the three components (see Section~4.2).

\medskip 
\centerline{\bf -- TABLE~5: Radio parameters of detected sources --}

From a quick comparison with the integrated NVSS 1.4\,GHz flux
densities, it is found that in many cases the high-resolution
observations must have resolved a substantial fraction of the emission. 
This can be quantified by examination of the spectral index values. 
Table~6 lists spectral indices $\alpha^{8.4}_{1.4}$ ($S_{\nu } \propto
\nu ^{-\alpha}$) (column~5), obtained by combining the NVSS flux 
densities (column~2)
and the integrated 8.4\,GHz flux densities (column~3) taken from Table~5.  
Due to the fact that the NVSS beam
(45\arcsec) exceeds the 8.4\,GHz beam (7\arcsec--10\arcsec) by a
substantial factor, these spectral index values should be considered
as upper limits. It indeed
appears that in most of these cases the radio emission was documented to
be extended over tens of arcseconds. 
We also compile calculated NVSS radio luminosities (column~4) for the 
37 galaxies under consideration.  
Again, for NGC\,3627 quoted values refer to
the compact core except when in parentheses, in which case the values refer 
to all three source components.

\medskip \centerline{\bf -- TABLE~6: summary  --}

Most sample galaxies are seen to display steep spectral indices, as
commonly found for star-forming late-type galaxies (Condon 1992).  The
following objects have flat ($\alpha \leq 0.6$) spectra: NGC\,4552,
NGC\,5354, NGC\,5838 and NGC\,5846 and the H\,II nucleus in NGC\,4424. 
We will return to this issue in the next Section. 

\section {Discussion}

Our VLA imaging observations have yielded X-band detections of 27 of the
37 sample galaxies.  These 27 include NGC\,4643 and NGC\,5012, which
were marginally detected.  As for the non-detections, they are equally
distributed over the transition LINERs and the H\,II nuclei.  It appears
that non-detection at X-band, at a resolution of
2\arcsecpoint5--12\arcsec, correlates with 1.4\,GHz (NVSS) weakness
and/or low surface brightness.  Table~6 indicates that the NVSS radio
luminosity distributions of the subsamples are comparable.  The
integrated 1.4\,GHz radio luminosities imply that the sample sources
span the usual luminosity range for nearby galaxies (10$^{26}$ --
10$^{30.5}$ erg s$^{-1}$ Hz$^{-1}$; see, e.g., Condon 1987, 1992). 
With the exception of NGC\,4424 (see Section~4.2), none of the H\,II 
nuclei display compact nuclear emission and/or a flat radio spectrum.
This is in strong contrast to the transition LINER galaxies.

Purely based on the radio morphology, the transition LINER objects can
be divided into two categories.  The first is made up of galaxies
displaying extended, steep-spectrum, low surface brightness radio
emission, usually tracing the optical isophotes of the host galaxy.  The
second category refers to objects that, in addition to extended
emission, also show compact nuclear radio emission. The first category displays
radio morphologies that are consistent with their being due to
large-scale star formation; in several cases the radio morphology is
seen to trace the H\,II regions in the host.  However, we stress that
the presence of a very weak nuclear component in these objects may still 
be masked by the more dominant radio emission from star-forming regions. 
The second category, which we will refer to in the following as AGN
candidates, is comprised of NGC\,3627, NGC\,4552, NGC\,5354,
NGC\,5838, and NGC\,5846.  The last four of these are characterized with
a flat radio spectral index between 1.4 and 8.4 GHz (see end of Section~4.3). 
Judging from high-resolution observations which isolate the nuclear
emission, the compact nucleus of NGC\,3627 also displays a flat radio
spectrum (see Section~4.2). These AGN candidates are on average more
luminous than the non-AGN and, as we shall see below, their hosts
are early rather than late-type galaxies. While at first sight the H\,II nucleus
NGC\,4424 also belongs in this AGN candidate class, the radio image (Fig.~3d)
shows it to be resolved. Higher resolution observations
recently carried out by us indeed resolve all the nuclear emission in 
NGC\,4424 (see Section~4.2), in contrast to the AGN candidates 
from the transition LINER sample. As such there is a clear separation between
the non-AGN and AGN candidate classes.  

Apparently, galaxies with composite LINER and star-formation spectra
separate out into objects with and without a clear signature of an
AGN, that is to say, a compact, flat-spectrum radio source.  The prime
question to address, of course, is whether the optical spectra of
these classes differ in any way.  Deferring the full analysis of this
issue to another paper, we here just note that the class of AGN
candidates indeed displays somewhat stronger [N~II] and [S~II] lines
in their optical spectra.  As we will demonstrate in the forthcoming
paper in this series, the behavior of the so called $u$-parameter,
which measures the radio/far-IR ratio, also supports the weak AGN
classification.  However, like in the case of Seyfert galaxies showing
a $\sim$30\% incidence rate of compact radio cores (e.g., Norris et
al.  1990), there is not a one-to-one correspondence between compact
radio emission and nuclear activity.  We conclude that the radio
properties permit to isolate AGN candidates among the sample of
transition LINERs, although the absence of radio cores cannot be used to 
argue against their being AGN.  These conclusions strongly support the 
hypothesis which resulted from the VLA observations of transition LINER
NGC\,7331 by Cowan et al.  (1994).

Inspection of Table~6 readily shows that these AGN candidates are
hosted by rather early-type galaxies (E--Sb).  The relevant radio core
luminosities L$_{\rm 8.4\,GHz}$ range from $\sim$ 10$^{26.2}$ --
10$^{28.5}$ erg s$^{-1}$ Hz$^{-1}$, which is several orders of
magnitude less than the core luminosities in FR\,I or FR\,II type
(Fanaroff \& Riley 1974) radio galaxies and quasars, and in the range
of the weakest radio cores in Seyfert galaxies (Giuricin et al.
1996).  This, then, implies that weak LINER AGN preferentially occur
in bulge-dominated hosts, not necessarily just in ellipticals,
consistent with statistical results from other lines of evidence (Ho
et al.  1997b; Ho 1999b).  This is in agreement with the case of
Seyfert galaxies, for which the radio core emission was also found to
be stronger in early-type hosts (Giuricin et al.  1996).  As such,
these weak AGN differ from the more powerful FR\,I radio galaxies,
which are commonly associated with elliptical galaxies, not seldom
brightest cluster galaxies (Zirbel \& Baum 1995).  They exceed the
very weak nuclear source in the nearby transition LINER NGC\,7331
(Cowan et al.  1994) by a factor of $\sim 15$, and are comparable in
strength to the well-known variable radio core of the LINER nucleus in
M\,81 (e.g., Ho et al.  1999).  Hence, it must be concluded that at least some 
weak LINER AGN reveal themselves by low-power radio cores.  This supports
the analysis of Ho (1999a), who proposed that the weak nuclear radio
sources (radio power 10$^{26}$ -- 10$^{29}$ erg s$^{-1}$ Hz$^{-1}$) in
nearby elliptical and S0 galaxies are the low-luminosity counterparts
of more powerful AGN.

Finally, it is intriguing that the transition LINER in the elliptical
galaxy NGC\,6482 remained undetected in our observations (as well as in
the NVSS).  Given that NGC 6482 is the most distant object in our
sample, our non-detection would still allow a $\leq 10^{27}$ erg
s$^{-1}$ Hz$^{-1}$ compact radio source to reside in this object. 

Full analysis of our sample, including optical emission-line and infrared
luminosities, will be presented in forthcoming papers.  To address the
physical origin behind the findings presented here will be the challenge
for future work. 

\section{Conclusions}

A sample of composite LINER/H\,II galaxies and pure H\,II nuclei was
studied using the VLA (C-array) at 8.4\,GHz. On the basis of their
radio morphological properties, the composite sources can be divided
into objects with low surface brightness emission confined mainly to
the plane of the galaxy and objects displaying compact nuclear
components. The former
objects are similar in radio morphology to the H\,II nuclei in our
sample, whereas the latter show morphologies consistent with AGN. All
five of the LINER AGN are hosted by bulge-dominated galaxies (Hubble
type E--Sb), and four of them show flat spectral indices between 8.4 and
1.4\,GHz. In terms of radio luminosity, the present LINER AGN populate
the range of low core radio luminosities, which implies that classical AGN 
of low luminosity exist in a wide range of galaxy types.

\acknowledgments

M.~E.~F.  is supported by grant PRAXIS XXI/BD/15830/98 from the 
Funda\c c\~ao para a Ci\^encia e Tecnologia, Minist\'erio da Ci\^encia e
Tecnologia, Portugal.  P.~D.~B.  acknowledges a visitor's grant from the
Space Telescope Science Institute, where a large part of this paper
could be written.  L.~C.~H.  is partly funded by NASA grant NAG 5-3556,
and by NASA grants GO-06837.01-95A and AR-07527.02-96A from the Space
Telescope Science Institute (operated by AURA, Inc., under NASA contract
NAS5-26555).  We want to thank Neil Nagar for providing us with 
some of his preliminary results.

This research was supported in part by the European
Commission TMR Programme, Research Network Contract ERBFMRXCT96-0034
``CERES.'' We made extensive use of the APM (Automatic Plate Measuring)
Facility, run by the Institute of Astronomy in Cambridge, the STScI DSS
(Digitized Sky Survey), produced under US government grant NAGW -- 2166,
and NED (NASA/IPAC Extragalactic Database), which is operated by the Jet
Propulsion Laboratory, California Institute of Technology, under
contract with NASA.

\clearpage 

\begin{table}
\footnotesize
\begin{center}
\begin{tabular}[h]{l c c c c c c r}
\tableline
\tableline

Galaxy & Other & R.A.(J2000) & Dec(J2000) & D &  & Hubble Type & Reference \\
 & name & $^{h}$ $^{m}$ $^{s}$ & $^{\circ}$ $^{'}$ $^{''}$ & Mpc & & & \\
\tableline 
IC\,520 &  & 08 53 42.3 & +73 29 29 & 47.0 & & SAB(rs)ab? & 26\\
NGC\,2541 & & 08 14 40.2  &  +49 03 42 & 10.6 & & SA(s)cd & 26,41\\
NGC\,2985 & & 09 50 20.9 & +72 16 44 & 22.4 & & (R')SA(rs)ab & 1,2,5,26,41,42 \\
NGC\,3627 & M\,66 & 11 20 14.9 & +12 59 21 & 6.6 & & SAB(s)b & 1,5,6,11,13,14\\
          &       &            &           &        &       &         & 17,24,26,28,41,44\\
NGC\,3628 & & 11 20 16.2 & +13 35 22 & 7.7 & & SAb pec spin & 1,4,5,11,13,14,16\\
          &&&&&&&                                                      17,18,19,24,26,30\\
         &&&&&&&                                                       34,36,38,41,44,45\\
NGC\,3675 & & 11 26 08.0 & +43 34 58 & 12.8 & & SA(s)b & 1,3,17,25,26,41\\
      &&&&&&&                                                   42,47\\
NGC\,3681  & & 11 26 29.4 & +16 51 51 & 24.2 & & SAB(r)bc & 1\\
NGC\,4013 & & 11 58 31.1 &  +43 56 50 & 17.0 & & SAb spin & 1,3,15,26,41\\
NGC\,4321 & M\,100& 12 22 54.8 &  +15 49 20 & 16.8 & & SAB(s)bc & 1,5,7,11,13,14,17\\
                                               &&&&&&&                   18,20,22,24,26,27\\
                                                &&&&&&&                  35,36,41,47\\
NGC\,4414 & & 12 26 27.2 & +31 13 24 & 16.8 & & SA(rs)c? & 1,3,5,11,13,14\\
                                       &&&&&&&                     17,18,24,26,31,36\\
NGC\,4552 &  M\,89 & 12 35 39.9 & +12 33 25 & 9.7 & & E & 1,13,14,17,22,29\\
                         &&&&&&&                                  37,38,39,41,43,44\\
NGC\,4643 & & 12 43 20.2 &  +01 58 41 & 16.8& & SB(rs)0/a & 21,24,41\\
NGC\,4713 & & 12 49 57.8 &  +05 18 39 & 25.7 & & SAB(rs)d & 1\\
NGC\,4826 & M\,64 &  12 56 44.2 & +21 41 05 & 17.9 & & (R)SA(rs)ab & 1,2,8,13,14,17,24,25\\
                                        &&&&&&&                              26,36,40,41,44,46\\
NGC\,5012 & &13 11 36.8 & +22 54 56 & 4.1 & & SAB(rs)c & 1\\
NGC\,5354 && 13 53 26.7 & +40 18 09 & 32.8 & & SA0 spin & 1,3,12,17,32,41,44\\
NGC\,5656 && 14 30 25.5 & +35 19 17 & 42.6 & & SAab & 1\\
NGC\,5678 && 14 32 05.2 & +57 55 23 & 35.6 & & SAB(rs)b & 1,13,14,17,18,31\\
NGC\,5838 && 15 05 26.5 & +02 06 01 & 28.5 & & SA0$^{-}$ & 1\\
NGC\,5846 && 15 06 29.2 & +01 36 21 & 28.5 & & E0$^{-}$ & 1,9,10,12,41\\
NGC\,5879 && 15 09 47.1 & +57 00 05 & 16.8 & & SA(rs)bc? & 1,41\\
NGC\,5921 && 15 21 56.3 & +05 04 11 & 25.2 & & SB(r)bc & 1,24,26,41\\
NGC\,6384 && 17 32 24.5 & +07 03 37 & 26.6 & & SAB(r)bc & 5,17,24,26,41\\
NGC\,6482 && 17 51 48.9 & +23 04 20 & 52.3 & & E: & \nodata \\
NGC\,6503 && 17 49 27.5 & +70 08 41 & 6.1 & & SA(s)cd & 1,5,26,41,42\\

\tableline
\end{tabular}
\end{center}
\setcounter{table}{0}
\caption{Sample sources -- the 25 transition LINERs.}
\end{table}

\clearpage

\begin{table}
\footnotesize
\begin{center}
\begin{tabular}[h]{l c c c c c c r}

\tableline
\tableline

Galaxy & Other & R.A.(J2000) & Dec(J2000) & D & & Hubble Type & Reference \\
 & name & $^{h}$ $^{m}$ $^{s}$ & $^{\circ}$ $^{'}$ $^{''}$ & Mpc & & & \\
\tableline
NGC\,3593 & & 11 14 36.0 & +12 49 06 & 5.5 & & SA(S)0/a: & 1,5,11,13,14,17\\
                       & &            &           &        &      &           & 18,24,26,33,41\\
NGC\,3684  & & 11 27 11.1 & +17 01 49 & 23.4 & & SA(rs)bc & 1\\
NGC\,4100 & & 12 06 08.1 & +49 34 59 & 17.0 & & (R')SA(rs)bc & 1,3,5,26,41,42\\
NGC\,4217 & & 12 15 50.7 & +47 05 37 & 17.0 & & SAb spin & 1,3,5,11,13,14\\
    &&&&&&&                                                                     15,17,26,41\\
NGC\,4245 & & 12 17 36.7 & +29 36 29 & 9.7 & & SB(r)0/a & \nodata\\
NGC\,4369 & Mrk\,439 & 12 24 36.1 & +39 22 58 & 21.6 & & (R)SA(rs)a & 1,3,18,41\\
NGC\,4405 & IC\,788 & 12 26 07.1 & +16 10 51 & 31.5 & & SA(rs)0: & 1\\
NGC\,4424 & & 12 27 11.4 & +09 25 15 & 16.8 & & SA(s)a: & 1\\
NGC\,4470 & & 12 29 37.9 & +07 49 25 & 31.4 & & Sa ? & 1\\
NGC\,4710 & & 12 49 38.9 &  +15 09 55 & 16.8 & & SA(r)0+? spin & 1,12,15,18,32,36\\
                                                         &&&&&&&       41,43\\
NGC\,4800 && 12 54 38.0 & +46 31 52 & 15.2 & & SA(rs)b & 1,3,41\\
NGC\,4845 & & 12 58 01.3 & +01 34 30& 15.6 & & SA(s)ab spin & 1,18\\
\tableline
 
\end{tabular}
\end{center}
\setcounter{table}{0}
\caption{Sample sources -- the 12 H\,II nuclei (cont.).}
\end{table}

\begin{table}
\small
\begin{center}
\begin{tabular}[h]{ll|ll}
\tableline
\tableline
No. & Reference & No. & Reference \\
\tableline

1 & Condon \etal 1998a & 25 & Gioia \& Fabbiano 1987 \\
2 & Condon \etal 1998b & 26 & Condon 1987 \\
3 & Becker \etal 1995 & 27 & Urbanik \etal 1986\\
4 & Dumke \etal 1995 & 28 & Urbanik \etal 1985\\
5 & Niklas \etal 1995 & 29 & Wrobel \& Heeschen 1984\\
6 & Saikia \etal 1994 &  30 & Schlickeiser et al. 1984\\
7 & Collison \etal 1994 & 31 & Condon  1983\\
8 & Turner \& Ho 1994 & 32 & Hummel \&  Kotanyi 1982\\
9 & Slee \etal 1994 &  33 & Jenkins \etal 1982 \\
10 & M\"ollenhoff \etal 1992  & 34 & Condon \etal 1982\\
11 & White \& Becker 1992 &  35 & Weiler \etal 1981\\
12 & Wrobel \& Heeschen 1991 & 36 & van der Hulst \etal 1981\\
13 & Becker \etal  1991 &  37 & van Breugel \etal 1981\\
14 & Gregory \& Condon 1991 & 38 & Jones \etal 1981a\\
15 & Hummel \etal 1991 & 39 & Jones \etal 1981b\\
16 & Reuter \etal 1991 & 40 & Klein \& Emerson 1981\\
17 & Condon \etal 1991 & 41 & Hummel \etal 1980\\
18 & Condon \etal 1990 & 42 & Heckman \etal 1980 \\
19 & Carral \etal 1990 & 43 & Dressel \& Condon 1978 \\
20 & Vila \etal 1990 & 44 & Sramek 1975a \\
21 & Fabbiano \etal 1989 & 45 & Haynes \& Sramek 1975\\
22 & Turner \etal 1988 &  46 & van der Kruit 1973a\\
23 & Condon \&  Broderick 1988 & 47 & van der Kruit 1973b\\ 
24 & Hummel \etal 1987 &  & \\

\tableline 
\end{tabular}
\end{center}
\setcounter{table}{1}
\caption{References for the radio data (see Table~1).}
\end{table}

\clearpage


\begin{table}
\footnotesize
\begin{center}
\begin{tabular}[h]{lccrcr}
\tableline
\tableline

Galaxy & Taper & Beamsize & PA & $rms$ & Fig. No. \\
 & K$\lambda $ & arcsec$^2$ & $^{\circ }$ & mJy/beam & \\ 
\tableline 
NGC\,2985  & 0 & 2.98 $\times $ 2.15 & $-$7.19 & 0.08 & \\
                    & 20 & 9.30 $\times $ 7.20 & $-$84.28 & 0.09 & 1a\\
NGC\,3627 & 0 & 2.55 $\times $ 2.29 & $-$25.24 & 0.10 & \\
                   & 15 & 10.96 $\times $ 9.93 & $-$64.54 & 0.25 & 1c\\	   
NGC\,3628  & 0 & 2.54 $\times $ 2.30 & $-$14.68 & 0.45 & \\
                    & 20 & 8.78 $\times $ 7.48 & $-$76.82 & 0.80 & 1d\\
NGC\,3675 & 25 & 7.27 $\times $ 5.81 & 71.04 & 0.09 & 2a\\
NGC\,4013  & 0 & 2.52 $\times $ 2.26 & $-$19.39 & 0.07 & \\
                    & 30 & 5.90 $\times $ 5.26 & 77.17 & 0.09 & 2b\\
NGC\,4321  & 0 & 2.52 $\times $ 2.33 & $-$11.27 & 0.07 & \\
                    & 30 & 7.25 $\times $ 5.14 & $-$89.95 & 0.08 & 3a\\
NGC\,4414 & 0 & 2.43 $\times $ 2.32 & $-$37.18 & 0.08 &\\
                   & 10 & 18.70 $\times $ 17.70 & $-$62.41 & 0.15 & 3c\\
NGC\,4552 & 0 & 2.93 $\times $ 2.54 & 11.27 & 0.08 & 4b\\
NGC\,4643 & 15 & 12.04 $\times $ 11.13 & 52.57 & 0.07 & 4c\\
NGC\,4713 & 15 & 12.12 $\times $ 10.95 & 44.22 & 0.08 & 5a\\
NGC\,4826 & 0 & 2.76 $\times $ 2.29 & $-$6.23 & 0.08 & \\
                   & 20 & 9.62 $\times $ 8.26 & 62.33 & 0.10 & 5c\\
NGC\,5012 & 15 & 11.59 $\times $ 11.18 & 74.66 & 0.08 & 6a\\
NGC\,5354 & 0 & 2.84 $\times $ 2.37 & $-$42.75 & 0.06 & 6b \\
                 
NGC\,5656 & 20 & 9.54 $\times $ 8.39 & 75.33 & 0.08 & 6c\\
NGC\,5678 & 0 & 2.99 $\times $ 2.21 & $-$27.47 & 0.07 &\\
                   & 15 & 11.09 $\times $ 10.75 & $-$34.43 & 0.10 & 6d\\
NGC\,5838 & 0 & 3.13 $\times $ 2.43 & 1.94 & 0.07 & \\
                   & 35 & 5.62 $\times $ 5.23 & 51.83 & 0.07 & 7a\\
NGC\,5846 & 0 & 3.13 $\times $ 2.49 & 3.94 & 0.07 &\\
                   & 30 & 7.06 $\times $ 5.67 & 68.01 & 0.07 & 7b\\
NGC\,5921 & 20 & 7.12 $\times $ 5.63 & 67.8 & 0.06 & 7c \\
\tableline
\tableline

Galaxy & Taper & Beamsize & PA & $rms$ & Fig. No. \\
 & K$\lambda $ & arcsec$^2$ & $^{\circ }$ & mJy/beam & \\ 
\tableline

NGC\,3593 & 0 & 2.58 $\times $ 2.36 & $-$17.94 & 0.09 & \\ 
                   & 20 & 9.70 $\times $ 7.17 & $-$80.16 & 0.15 & 1b \\
NGC\,4100 & 0 & 2.61 $\times $ 2.35 & $-$29.79 & 0.07 & \\
                  & 20 & 7.83 $\times $ 7.40 & 84.07 & 0.08 & 2c \\
NGC\,4217 & 20 & 7.30 $\times $ 5.41 & 76.75 & 0.09 & 2d\\
NGC\,4369 & 0 & 2.48 $\times $ 2.34 & $-$14.62 & 0.07 & \\
                   & 30 & 6.94 $\times $ 5.51 & 88.71 & 0.10 & 3b \\
NGC\,4424 & 0 & 3.00 $\times $ 2.46 & 12.14 & 0.09 & \\
                   & 30 & 7.21 $\times $ 5.60 & 76.27 & 0.07 & 3d\\
NGC\,4470 & 15 & 9.95 $\times $ 7.99 & 59.70 & 0.07 & 4a\\
NGC\,4710  & 0 & 2.86 $\times $ 2.48 & 3.88 & 0.10  &\\
                   & 30 & 5.50 $\times $ 5.16 & 72.46 & 0.08 & 4d\\
NGC\,4800 & 30 & 5.28 $\times $ 5.15 & $-$82.56 & 0.10 & 5b\\
NGC\,4845 & 0 & 3.27 $\times $ 2.44 & 15.08 & 0.08 & \\
                   & 30 & 7.08 $\times $ 5.69 & 64.51 & 0.10 & 5d \\

\tableline
\end{tabular}
\end{center}
\setcounter{table}{2}
\caption {Map parameters of the detected sources. The H\,II nuclei appear
below the transition LINERs.}
\end{table}

\clearpage


\begin{table}
\footnotesize
\begin{center}
\begin{tabular}[h]{lcccccrr}
\tableline
\tableline

Field Source & R.A.(J2000) & Dec(J2000) & z & NVSS & F$_{\rm 8.4}^{\rm int}$ & $\alpha ^{\rm 8.4}_{\rm 1.4}$ & 
Reference \\
             & $^{h}$ $^{m}$ $^{s}$ &  $^{\circ }$ $^{'}$ $^{''}$ & & mJy &  mJy  & & \\ 
\tableline

NGC\,3593N & 11 14 37.20 & +12 52 15.0 & \nodata & 18.7 & 10.6 & 0.3 & \nodata\\
NGC\,3684SW   & 11 27 02.42 & +16 58 34.8 & \nodata & 41.4  & 18.1 & 0.5 & \nodata\\
NGC\,4013SE & 11 58 38.61 & +43 55 05.5 & \nodata & \nodata & 3.8 & \nodata & \nodata\\
SN\,1979C & 12 22 58.67 & +15 47 51.6 & 0.0052 & \nodata & 1.3 & \nodata  & Weiler \etal 1981 \\
LBQS\,1223+1626 & 12 25 59.09 & +16 10 21.2 & 1.9290 & 8.9 & 10.1 & $-$0.1 & Hewett \etal 1995 \\
NGC\,4424SE & 12 27 19.83 & +09 23 03.0 & \nodata & 11.0 & 13.3 &  $-$0.1 & \nodata \\
TXS\,1227+081 & 12 29 47.64 & +07 50 26.7 & \nodata & 110.9 & 16.0 & 1.1 & Douglas \etal 1996 \\
NGC\,4710E & 12 49 48.63 & +15 09 33.3 & \nodata & 4.1 & 4.8  & $-$0.1& \nodata \\ 
NGC\,5353 & 13 53 26.69 & +40 16 58.7 & 0.0077 & 41.0 & 26.7 & 0.2 & Hummel \& Kotanyi 1982\\
NGC\,5838S & 12 05 28.11 & +02 04 16.2 & \nodata  & 7.5 & 1.3 & 1.0 & \nodata \\                  
87GB\,1508+5714 & 15 10 02.97 & +57 02 43.6 & 4.3010 & 202.4 & 292.1 & $-$0.2 & Hook \etal 1995 \\
87GB\,1749+2302 & 17 51 49.10 & +23 01 26.7 & 0.7700 & 45.3 & 23.6 & 0.4  & Becker \etal 1991 \\
BL\,1749+701 & 17 48 32.87 & +70 05 52.5 & \nodata & 735.6 & \nodata & \nodata & Hughes \etal 1992 \\

\tableline 
\end{tabular}
\end{center}
\setcounter{table}{3}
\caption{The field source parameters. The integrated 8.4\,GHz flux densities 
(column~6) have been corrected for primary beam attenuation.}
\end{table}

\clearpage


\begin{table}
\footnotesize
\begin{center}
\begin{tabular}[h]{lccc cr}
\tableline
\tableline
Galaxy & Taper & F$_{\rm max}$ & R.A.(J2000) & Dec(J2000) & F$_{\rm int}$ \\
           & K$\lambda $ & mJy/beam & $^{h}$ $^{m}$ $^{s}$ &  $^{\circ }$ $^{'}$ $^{''}$ & mJy \\ 
\tableline 

NGC\,2985  & 0 & 1.07 & 09 50 22.1 & +72 16 44 & $\geq$1.9\\
           & 20 & 1.30 & 09 50 22.0 & +72 16 44 & $\geq$2.7 \\	
NGC\,3627 & 0 & 1.30 & 11 20 15.0 & +12 59 30 & 3.7 ($\geq$22.1) \\
                   & 15 & 2.99 & 11 20 15.1 & +12 59 31 & 3.9 ($\geq$25.7)\\
NGC\,3628  & 0 & 15.31 & 11 20 17.0 & +13 35 20 & 61.4 \\
                    & 20 & 41.07 & 11 20 17.0 & +13 35 19 & 69.1\\
NGC\,3675 & 25 & 0.37 & 11 26 08.7 & +43 35 01 &  $\geq$1.3 \\

NGC\,4013 & 0 &  1.08 & 11 58 31.4 & +43 56 51 &  $\geq$3.8 \\
                   & 30 & 2.28 & 11 58 31.4 & +43 56 51 &  $\geq$4.1 \\
NGC\,4321  & 0 & 0.34 & 12 22 55.4 & +15 49 21 & 7.5\\
                    & 30 & 1.64 & 12 22 55.3 & +15 49 21 & 15.7\\
NGC\,4414 & 0 & 0.65 & 12 26 27.5 & +31 13 40 & 2.3 \\
                   & 10 & 3.72 & 12 26 26.7 & +31 13 42 & 29.4\\
NGC\,4552  & 0 & 76.51 & 12 35 39.8 & +12 33 23 & 76.8 \\
   
NGC\,4643 & 15 & \nodata & \nodata & \nodata & $\sim$1 \\
NGC\,4713 & 15 & 0.32 & 12 49 58.4 & +05 18 39 &  $\geq$1.2\\
NGC\,4826 & 0 & 1.8 & 12 56 43.4 & +21 41 01 & 18.9\\
                   & 20 & 4.03 & 12 56 43.7 & +21 41 00 & 21.1\\
NGC\,5012 & 15 & \nodata& \nodata & \nodata & $\sim$1 \\
NGC\,5354 & 0 & 11.61 & 13 53 26.7 & +40 18 10 & 11.7 \\
                   
NGC\,5656 & 20 & 0.40 & 14 30 26.9 & +35 19 25 & $\geq$0.7\\
NGC\,5678 & 0 & 0.74 & 14 32 05.9 & +57 54 51 &  $\geq$2.0 \\
                   & 15 & 1.31 & 14 32 05.5 & +57 54 51 & $\geq$8.6\\

NGC\,5838 & 0 & 1.85 & 15 05 26.3 & +02 05 57 & 2.1\\
                    & 35 & 1.88 & 15 05 26.3 & +02 05 57 & 2.2 \\

NGC\,5846 & 0  & 6.03 & 15 06 29.3 & +01 36 21 & 6.4 \\
             & 30 & 6.36 & 15 06 29.3 & +01 36 21 & 7.1\\

NGC\,5921 & 20 & 0.46 & 15 21 56.4 & +05 04 14 &  $\geq$0.8 \\
\tableline
\tableline

Galaxy & Taper & F$_{\rm max}$ & R.A.(J2000) & Dec(J2000) & F$_{\rm int}$ \\
           & K$\lambda $ & mJy/beam & $^{h}$ $^{m}$ $^{s}$ &  $^{\circ }$ ' '' & mJy \\ 
\tableline 

NGC\,3593 & 0 & 1.82 & 11 14 36.4 & +12 49 06 & 16.4 \\ 
                   & 20 & 4.79 & 11 14 36.6 & +12 49 05 & 20.0 \\
NGC\,4100 & 0 & 0.93 & 12 06 08.6 & +49 34 58 &  $\geq$3.9 \\
                   & 20 & 3.58 & 12 06 08.4 & +49 34 59 &  $\geq$7.3 \\
NGC\,4217 & 20 & 0.55 & 12 15 51.0 & +47 05 29 &  $\geq$22.0 \\
NGC\,4369 & 0 & 0.31 & 12 24 36.3 & +39 22 56 & 1.3 \\
                   & 30 & 1.31 & 12 24 36.3 & +39 22 57 & 3.8 \\
NGC\,4424 & 0 & 0.60 & 12 27 11.2 & +09 25 17 & 2.0 \\
                   & 30 & 0.80 & 12 27 11.7 & +09 25 17 & 2.5 \\

NGC\,4470 & 15 & 0.38 & 12 29 37.8 & +07 49 24 & $\geq$2.6 \\
NGC\,4710 & 0 & 0.71 & 12 49 38.9 & +15 09 59 & $\geq$6.0 \\
                   & 30 & 1.64 & 12 49 39.0 & +15 09 58 & 6.1 \\
NGC\,4800 & 30 & 0.36 & 12 54 37.8 & +46 31 52 & $\geq$1.2 \\
NGC\,4845 & 0 & 1.76 & 12 58 01.2 & +01 34 32 & $\geq$9.9 \\
                   & 30 & 5.15 & 12 58 01.1 & +01 34 32 & 12.5 \\

\tableline 
\end{tabular}
\end{center}
\setcounter{table}{4}
\caption{The 8.4\,GHz radio parameters of the detected sources. The 
H\,II nuclei appear below the transition LINERs.}
\end{table}

\clearpage


\begin{table}
\footnotesize
\begin{center}
\begin{tabular}[h]{l c r c r c}
\tableline
\tableline

Galaxy & NVSS & F$_{\rm 8.4}^{\rm int}$ & Log L$_{\rm 1.4}^{\rm tot}$ & 
$\alpha_{\rm 1.4}^{\rm 8.4}$ & Hubble Type \\
 & mJy & mJy & erg s$^{-1}$ Hz$^{-1}$  & & \\
\tableline 
IC\,520 & \nodata & \nodata & \nodata & \nodata  & SAB(rs)ab?\\
NGC\,2541 & \nodata & \nodata & \nodata & \nodata & SA(s)cd \\
NGC\,2985 & 44.1 & 2.7 & 28.4 & 1.6 & (R')SA(rs)ab \\
NGC\,3627 & 324.9 & 3.9 (25.7) &  28.2 & 2.5 (1.5) & SAB(s)b \\
NGC\,3628 & 291.7 & 69.1 &  28.3 & 0.8 & SAb pec spin \\ 
NGC\,3675 & 48.9 & 1.3 & 27.9 & 2.0 & SA(s)b \\
NGC\,3681  & 4.2 &  \nodata & 27.5 &  \nodata & SAB(r)bc \\
NGC\,4013 & 40.5 & 4.1 &  28.2 & 1.3 & SAb spin \\
NGC\,4321 & 87.1 &  15.7 & 28.5 & 1.0 & SAB(s)bc\\
NGC\,4414 & 242.2 &  29.4 & 28.4 & 1.2 & SA(rs)c? \\
NGC\,4552 & 103.1  & 76.8 &  28.5 &  0.2 & E \\
NGC\,4643 & \nodata & $\sim$1 & \nodata &  \nodata & SB(rs)0/a \\
NGC\,4713 & 46.9 & 1.2 &  28.3 & 2.0 & SAB(rs)d \\
NGC\,4826 & 101.1 &  21.1 & 27.3 & 0.9 & (R)SA(rs)ab\\
NGC\,5012 & 31.4 & $\sim$1 &  28.8 & \nodata & SAB(rs)c \\
NGC\,5354 & 8.4 & 11.7 &  28.0  & $-$0.2 & SA0 spin \\
NGC\,5656 & 22.0 &  0.7 & 28.7  & 2.0 & SAab  \\
NGC\,5678 & 111.5 & 8.6 &  29.2  & 1.4 & SAB(rs)b \\
NGC\,5838 & 3.0 &  2.2 & 27.5  & 0.2 & SA0$^{-}$ \\
NGC\,5846 & 22.1 &  7.1 & 28.3 & 0.6 & E0$^{-}$ \\
NGC\,5879 & 21.1 &  \nodata & 27.9  & \nodata & SA(rs)bc?\\
NGC\,5921 & 24.2 & 0.8 &  28.3  & 1.9 & SB(r)bc\\
NGC\,6384 & \nodata & \nodata & \nodata & \nodata & SAB(r)bc\\
NGC\,6482 & \nodata & \nodata & \nodata & \nodata &  E:  \\
NGC\,6503 & 40.0 &  \nodata & 27.3 & \nodata & SA(s)cd\\

\tableline
\tableline
Galaxy & NVSS & F$_{\rm 8.4}^{\rm int}$ &   Log L$_{\rm 1.4}^{\rm tot}$& 
$\alpha_{\rm 1.4}^{\rm 8.4}$ & Hubble Type \\
 & mJy & mJy & erg s$^{-1}$  Hz$^{-1}$  & & \\
\tableline 
 
NGC\,3593 & 87.3 & 20.0 & 27.5 &  0.8 & SA(S)0/a: \\
NGC\,3684 & 15.9 &  \nodata & 28.0 & \nodata & SA(rs)bc  \\
NGC\,4100 & 50.3 &  7.3 & 28.2 & 1.1 & (R')SA(rs)bc \\
NGC\,4217 & 122.8 &  22.0 & 28.6 & 1.0&  SAb spin \\
NGC\,4245 & \nodata & \nodata & \nodata & \nodata & SB(r)0/a \\
NGC\,4369 & 24.3 & 3.8 &  28.1 & 1.0 & (R)SA(rs)a \\
NGC\,4405 & 4.5 &  \nodata & 27.7 & \nodata &  SA(rs)0: \\
NGC\,4424 & 4.5 &  2.5 & 27.2 & 0.3 & SA(s)a: \\
NGC\,4470 & 17.1 &  2.6 & 28.3 & 1.1 & Sa ? \\
NGC\,4710 & 19.3 &  6.1 & 27.8 &  0.7 & SA(r)0+? spin \\
NGC\,4800 & 23.5 &  1.2 & 27.8 &  1.7 & SA(rs)b  \\
NGC\,4845 & 43.8 &  12.5 & 28.1 & 0.7 &  SA(s)ab spin \\

\tableline
\end{tabular}
\end{center}
\setcounter{table}{5}
\caption{Summary of the detected sources. The H\,II nuclei appear
 below the transition LINERs. All tabulated spectral index values are upper
limits (see Section~4.3).}
\end{table}

\clearpage

\begin{figure}
\leavevmode
\centerline{
\epsfxsize=7.3cm
\epsffile{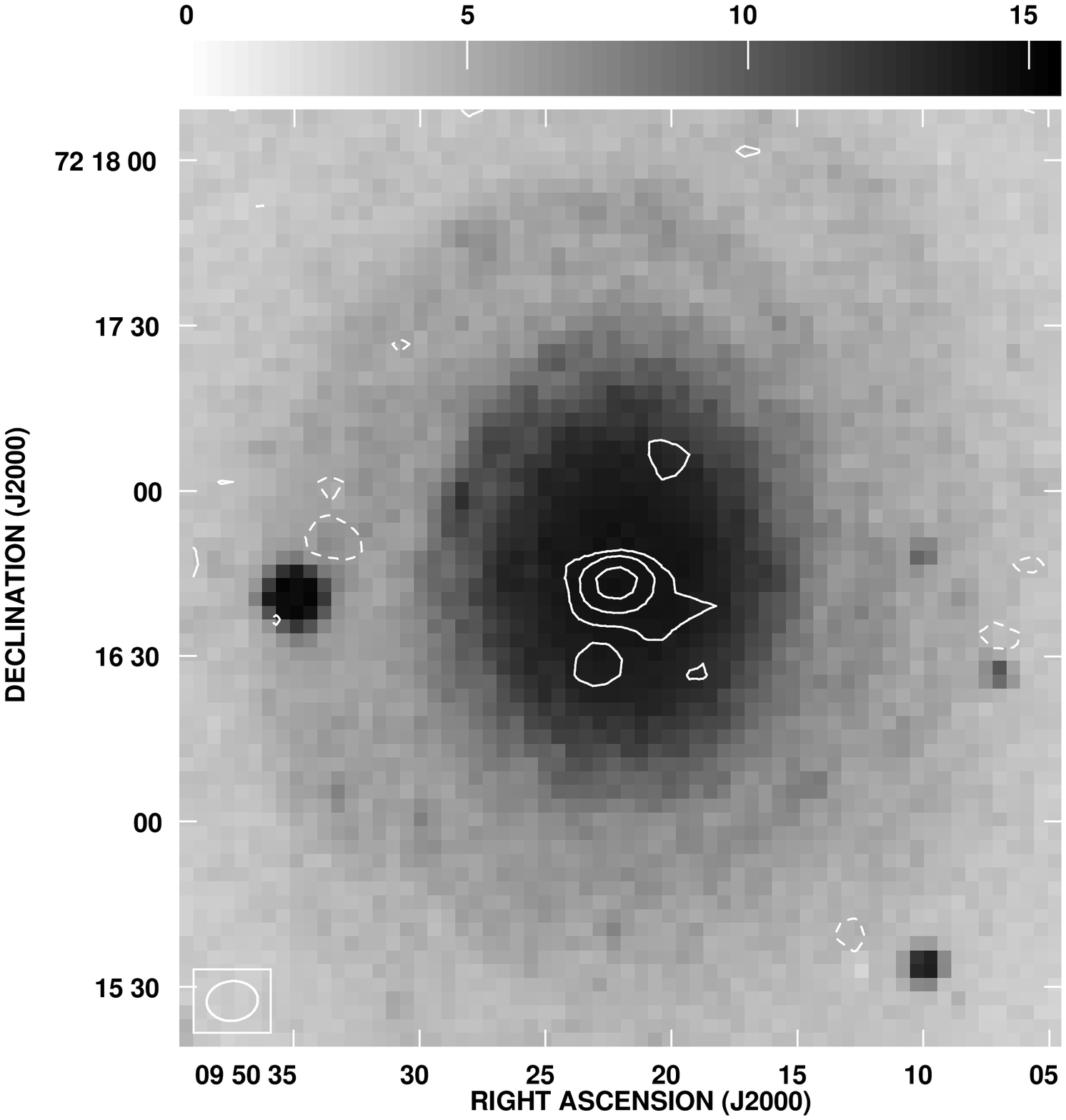}
(a) 
\epsfxsize=7.3cm
\raisebox{1.4cm}{
\epsffile{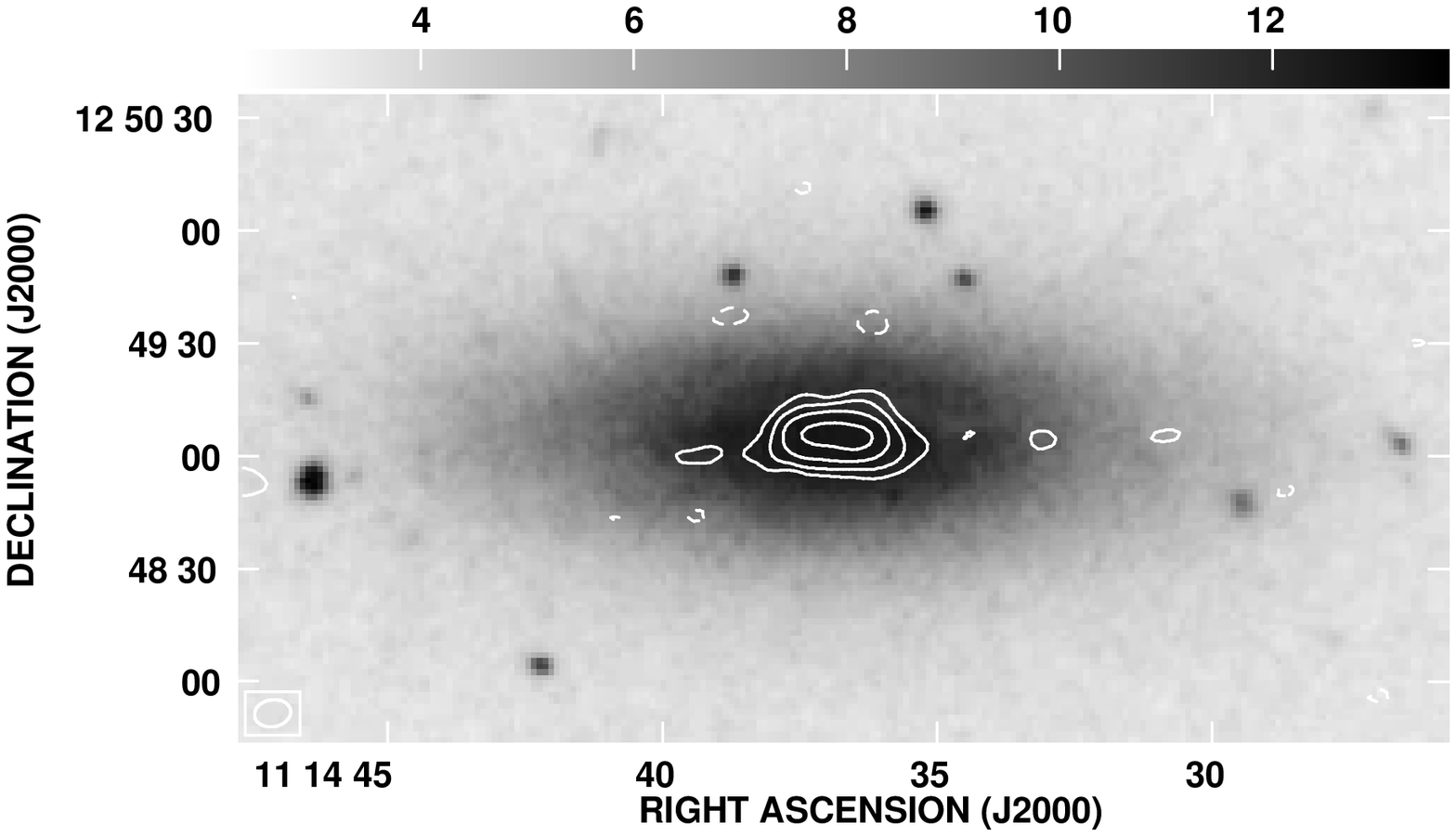}
(b)
}
}

\end{figure}

\begin{figure}
\figurenum{1}
\leavevmode
\centerline{
\epsfxsize=7.3cm
\epsffile{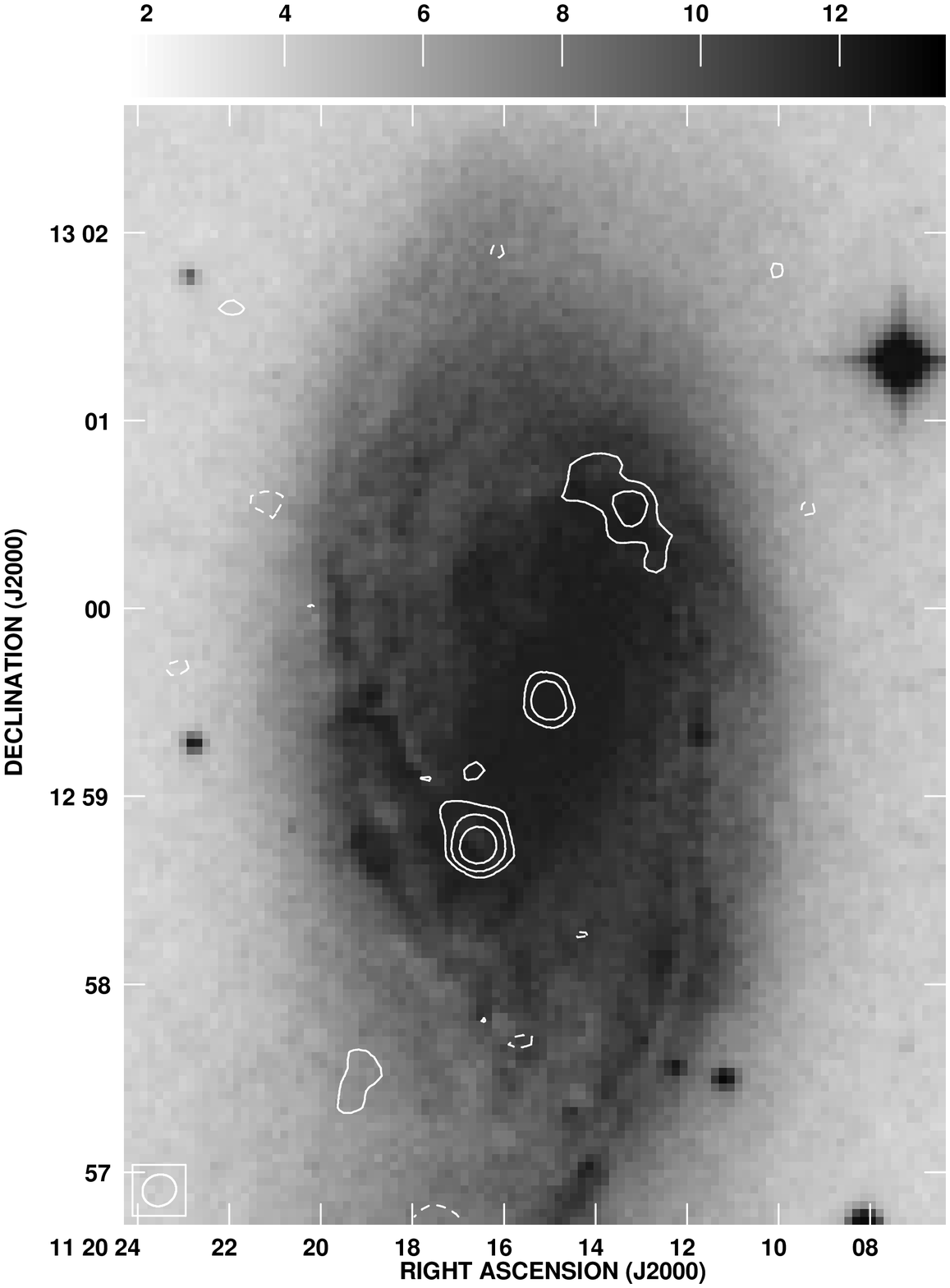}
(c)
\epsfxsize=7.3cm
\raisebox{1.4cm}{
\epsffile{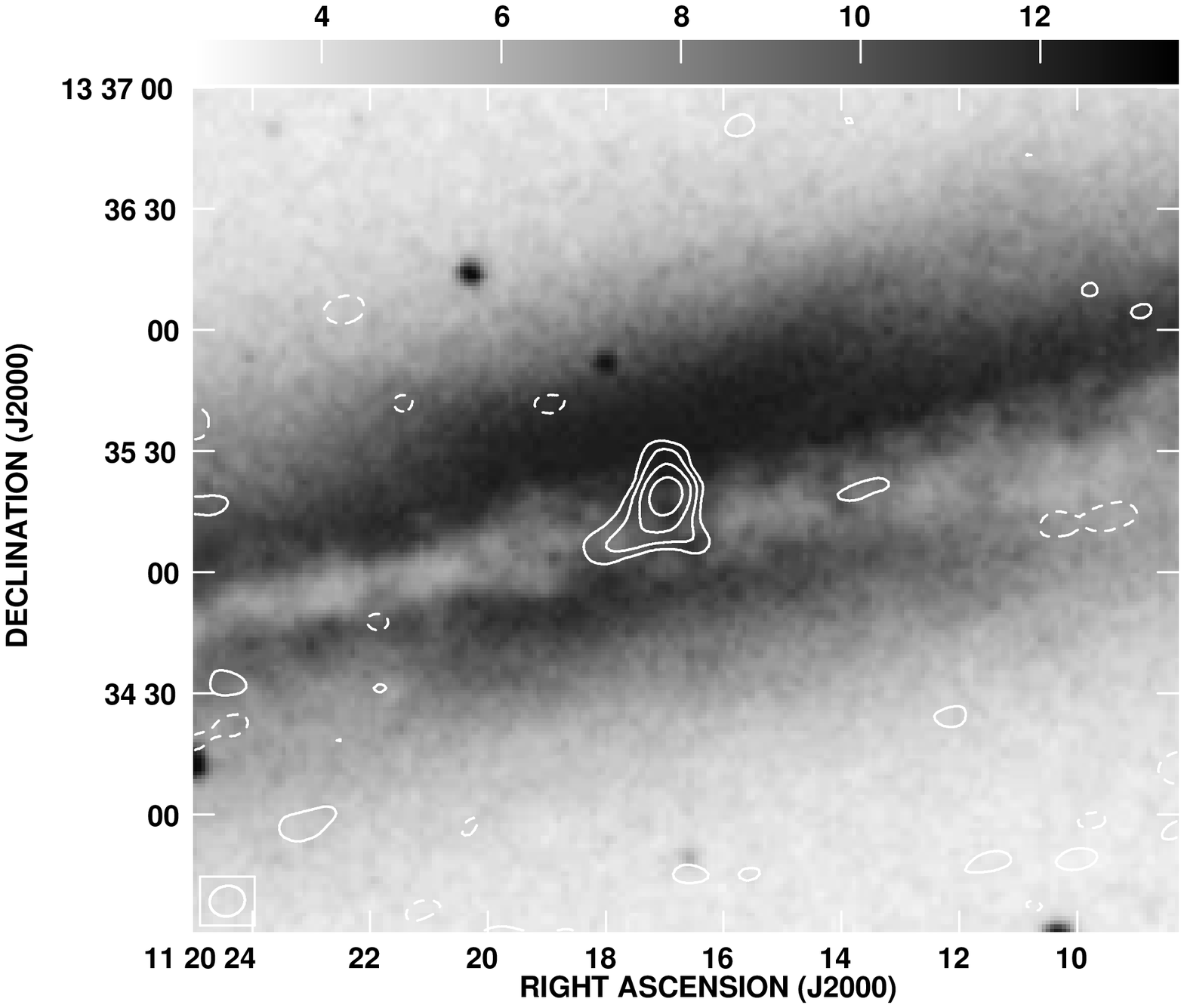}
(d)
}
}

\caption{Radio emission (contours) superimposed on optical images from 
the Digitized Sky Survey (greyscale). Contour levels are CLEV $\times$ 
(--3, 3, 6, 12, 24, 48, 96), where CLEV is the $rms$ noise level (see Table~3). 
The grey scale levels are arbitrary. The size of the restoring 
beam is given in parentheses after each object name (see Table~3).
{\bf (a)} NGC\,2985 (9\arcsecpoint30 $\times$ 7\arcsecpoint20), {\bf (b)} 
NGC\,3593 (9\arcsecpoint70 $\times$ 7\arcsecpoint17), {\bf (c)} NGC\,3627 (10\arcsecpoint96 $\times$ 9\arcsecpoint93) and {\bf (d)} NGC\,3628 
(8\arcsecpoint78 $\times$ 7\arcsecpoint48). NGC\,3593 is an H\,II nucleus. }

\end{figure}

\clearpage

\begin{figure}
\leavevmode
\centerline{
\epsfxsize=7.3cm
\epsffile{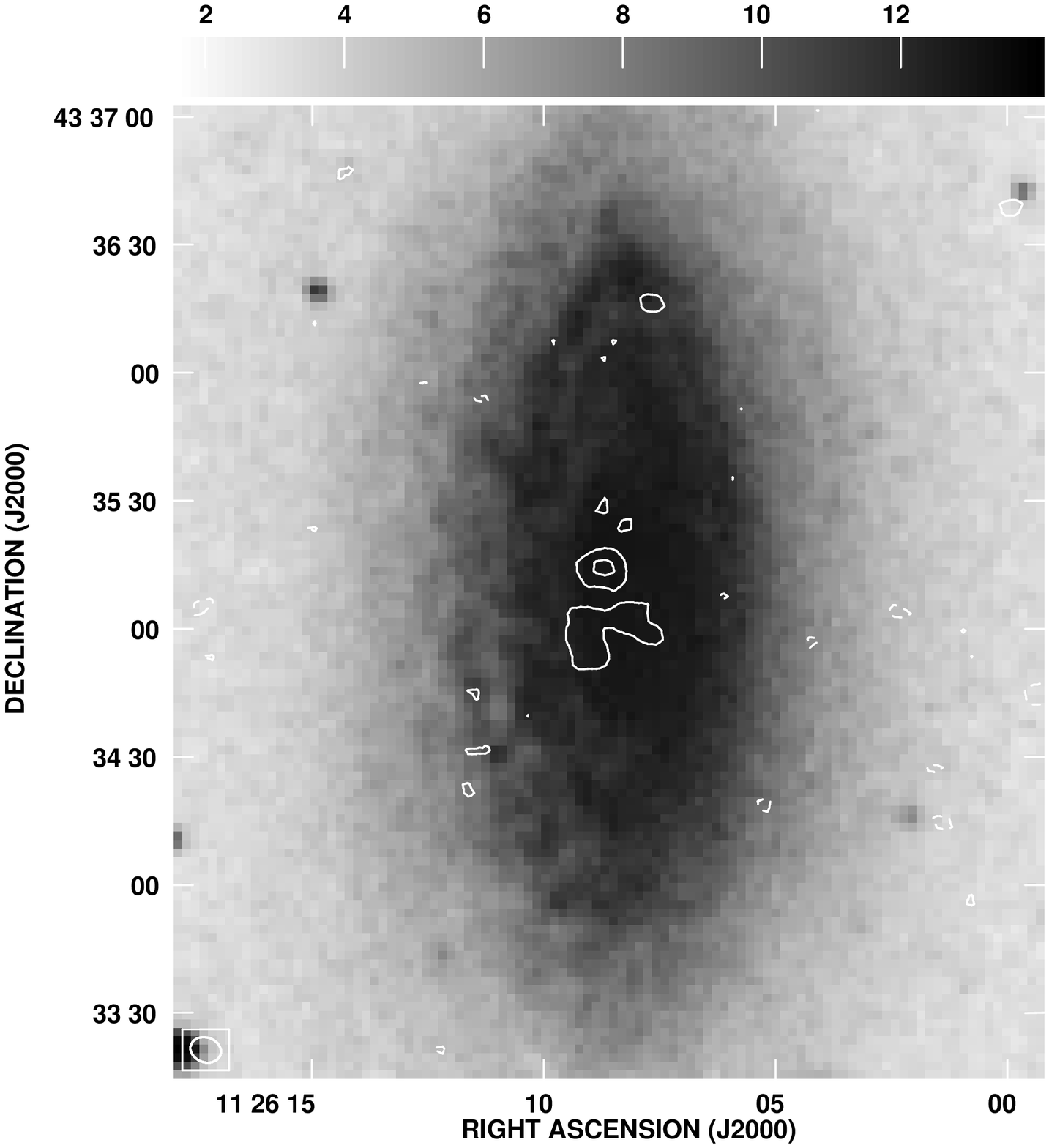}
(a)
\epsfxsize=7.3cm
\raisebox{1.8cm}{
\epsffile{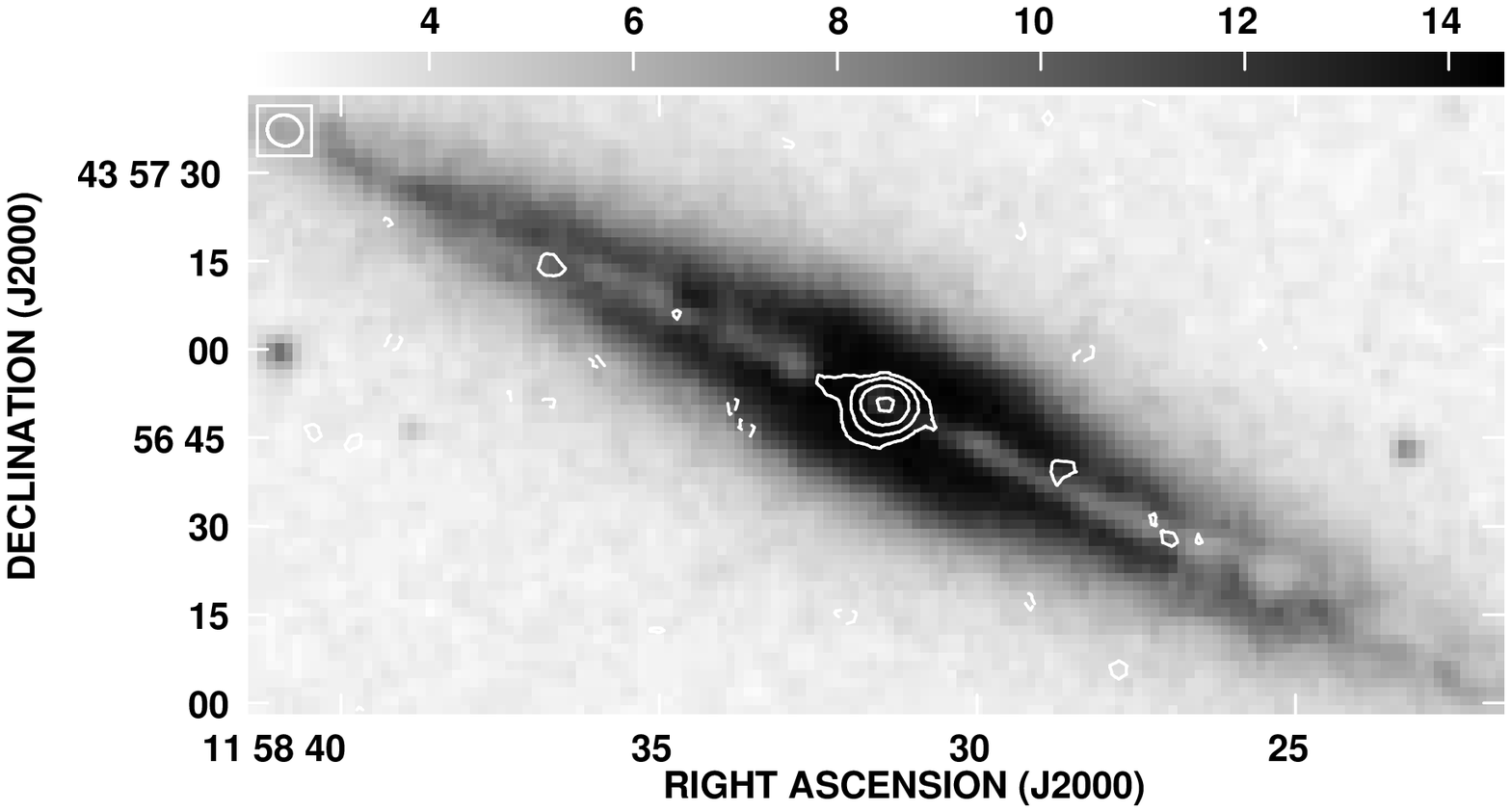}
(b)
}
}
\end{figure}

\begin{figure}
\figurenum{2}
\leavevmode
\centerline{
\epsfxsize=7.3cm
\epsffile{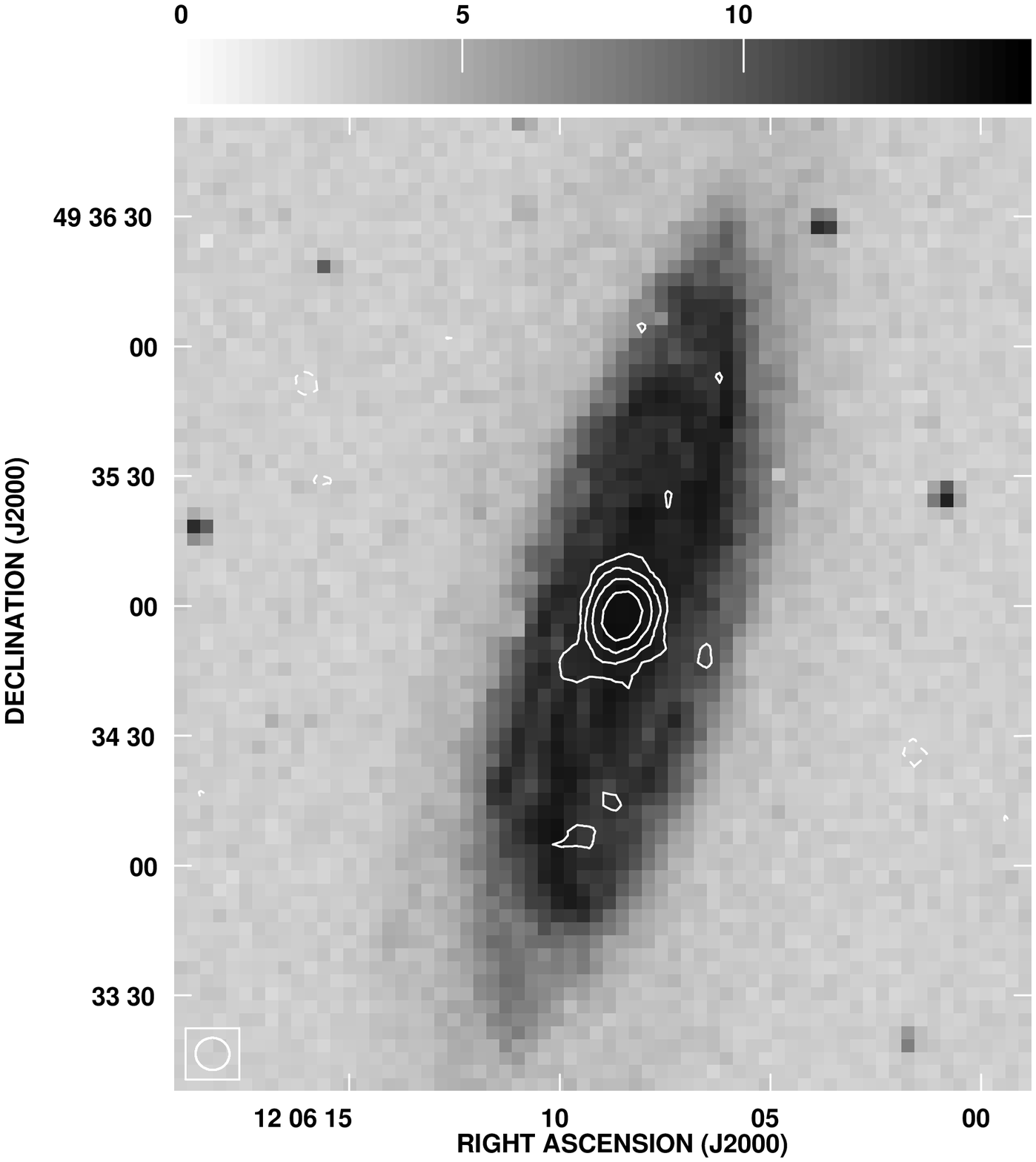}
(c)
\epsfxsize=7.3cm
\raisebox{1cm}{
\epsffile{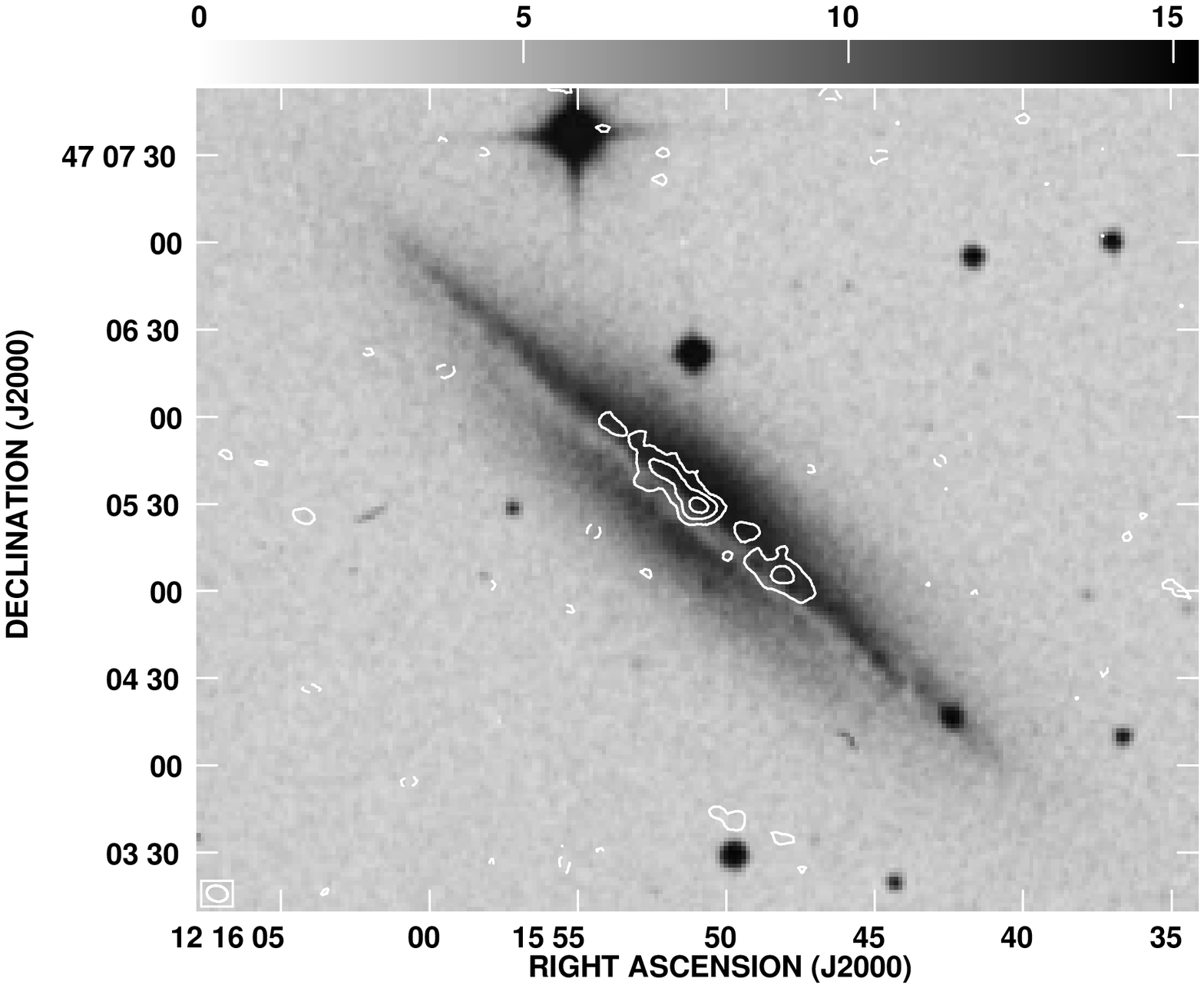}
(d)
}
}

\caption{As in Figure~1. {\bf (a)} NGC\,3675 (7\arcsecpoint27 $\times$ 5\arcsecpoint81), {\bf (b)} NGC\,4013
(5\arcsecpoint90 $\times$ 5\arcsecpoint26), {\bf (c)} NGC\,4100 (7\arcsecpoint83 $\times$ 7\arcsecpoint40) and {\bf (d)}
NGC\,4217 (7\arcsecpoint30 $\times$ 5\arcsecpoint41).  NGC\,4100 and NGC\,4217 are H\,II nuclei.}

\end{figure}

\clearpage

\begin{figure}
\leavevmode
\centerline{
\epsfxsize=7.3cm
\raisebox{0.1cm}{
\epsffile{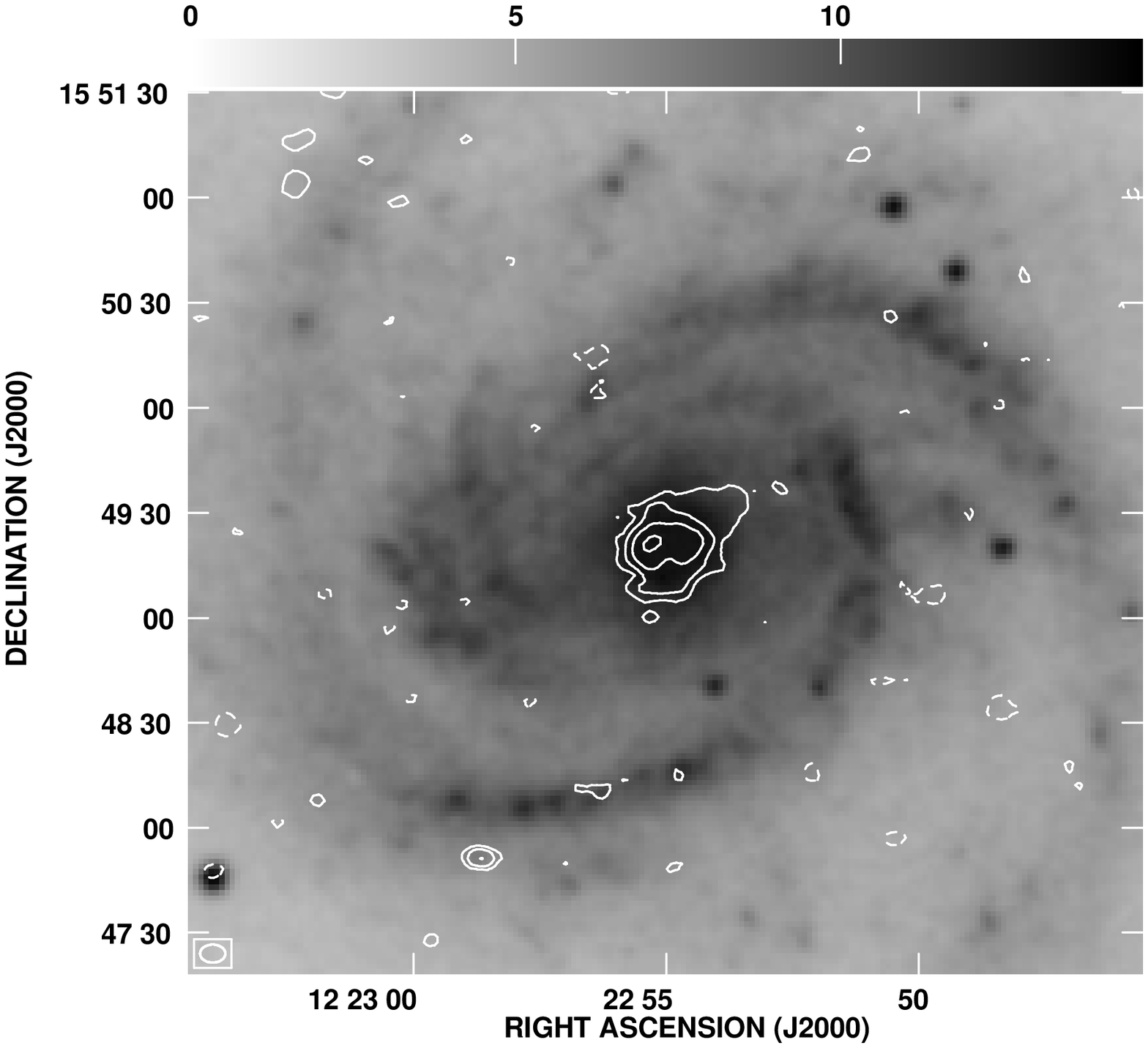}
(a)
}
\epsfxsize=7.3cm
\epsffile{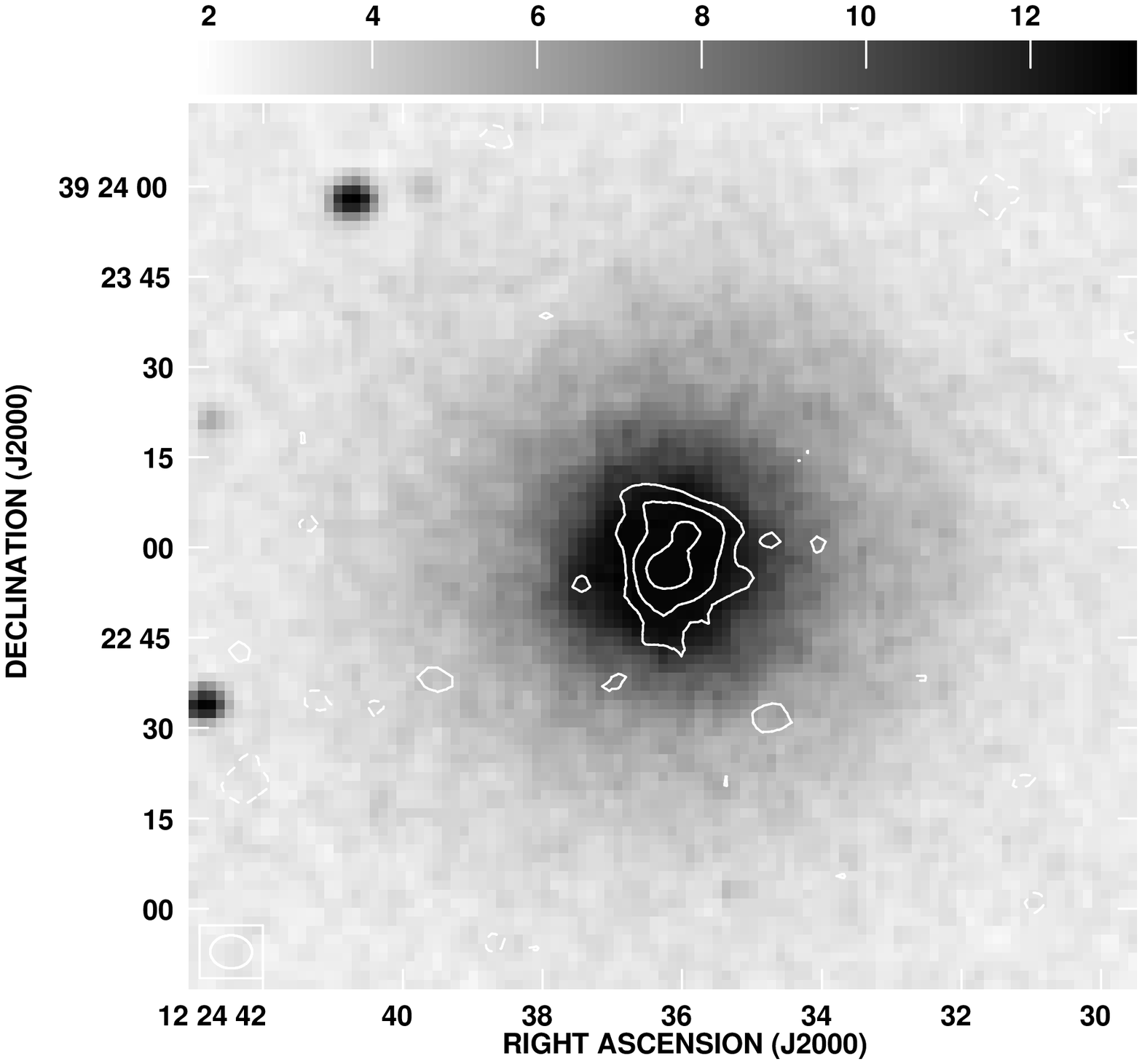}
(b)
}

\end{figure}

\begin{figure}
\figurenum{3}
\leavevmode
\centerline{
\epsfxsize=7.3cm
\epsffile{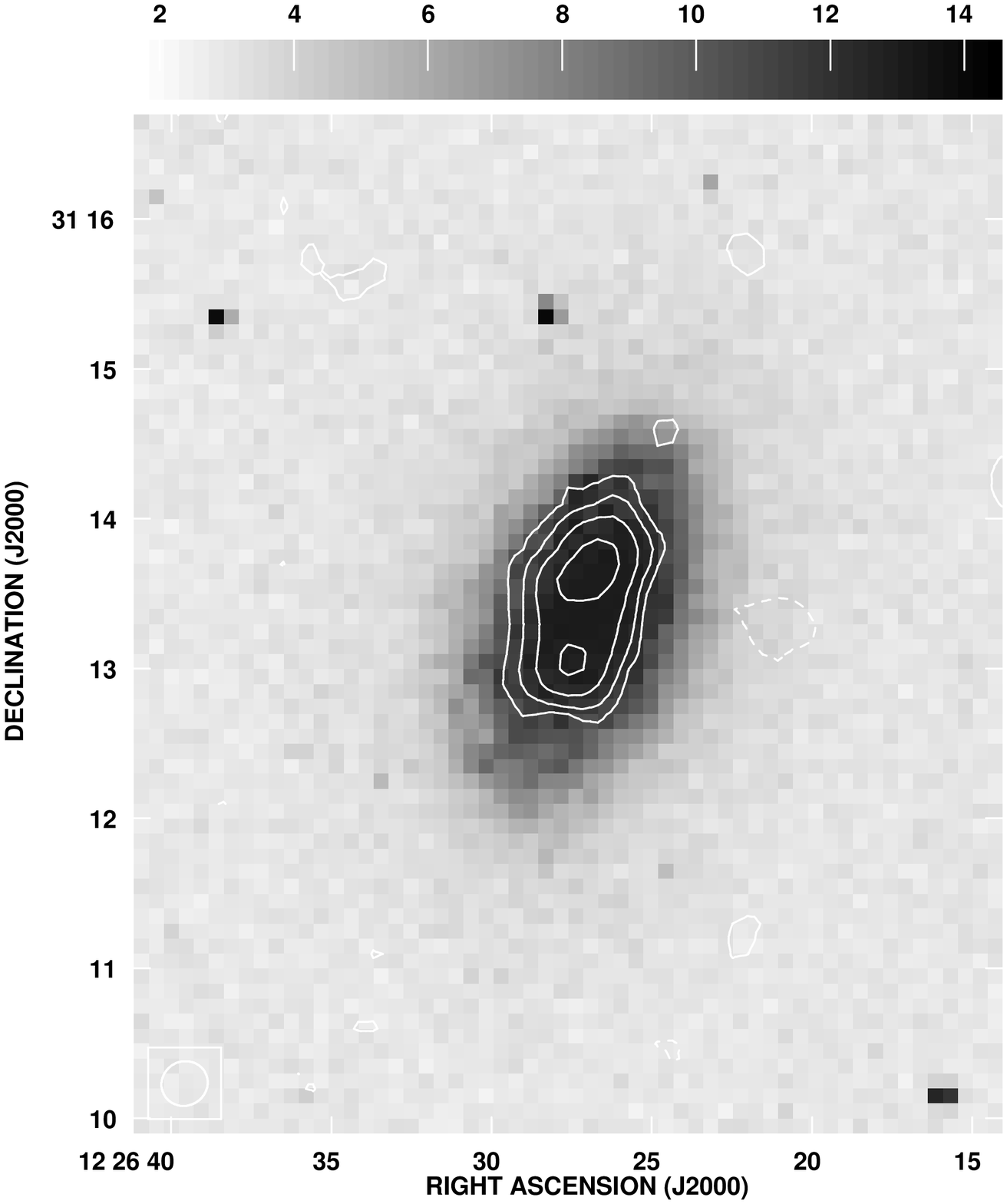}
(c)
\raisebox{1.6cm}{
\epsfxsize=7.3cm
\epsffile{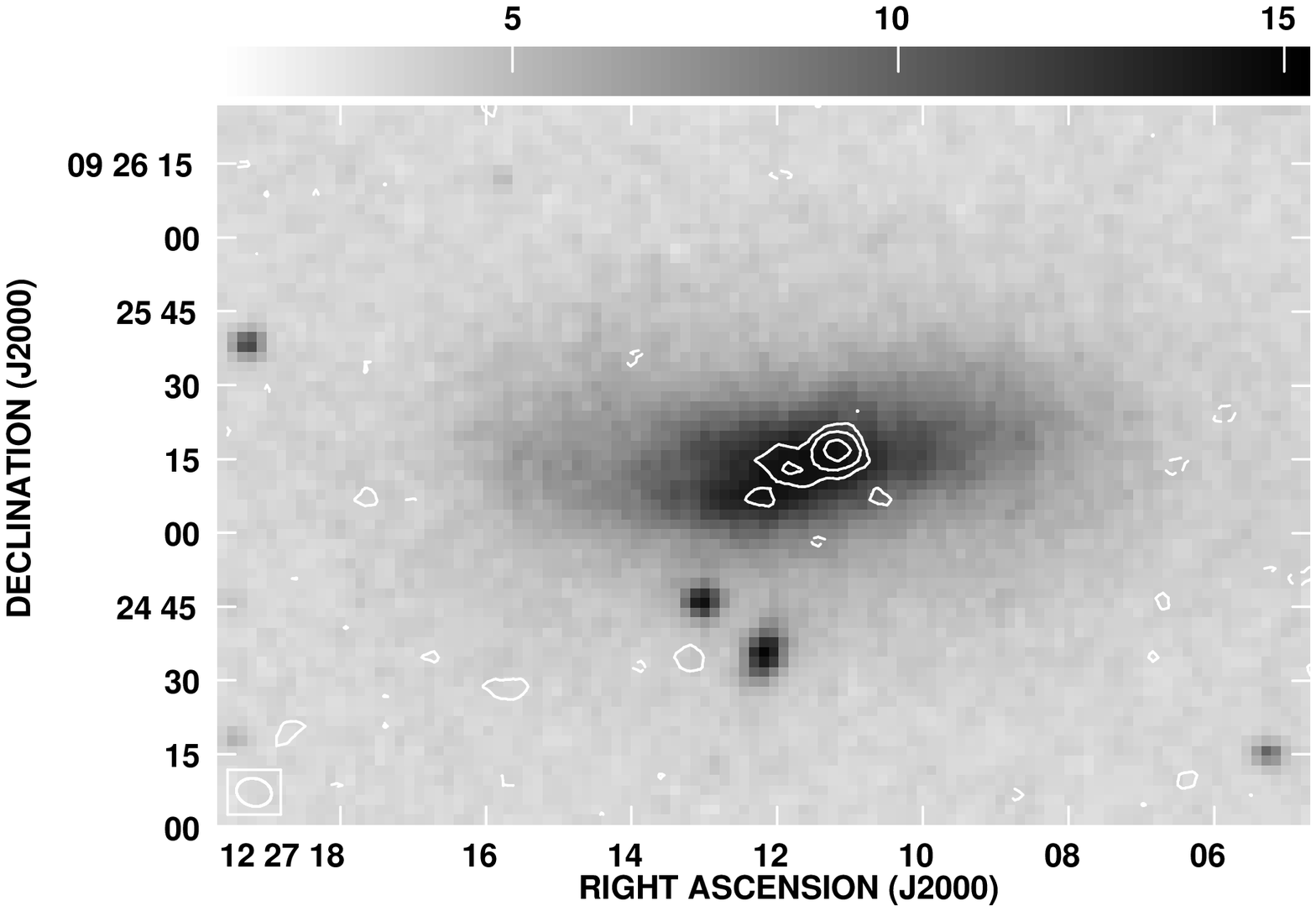}
(d)
}
}

\caption{As in Figure~1. {\bf (a)} NGC\,4321 (7\arcsecpoint25 $\times$ 5\arcsecpoint14), {\bf (b)} NGC\,4369
(6\arcsecpoint94 $\times$ 5\arcsecpoint51), {\bf (c)} NGC\,4414 (18\arcsecpoint17 $\times$ 17\arcsecpoint70) and {\bf (d)} NGC\,4424
(7\arcsecpoint21 $\times$ 5\arcsecpoint60).  NGC\,4369 and NGC\,4424 are H\,II nuclei.}

\end{figure}

\clearpage

\begin{figure}
\leavevmode
\centerline{
\epsfxsize=7.3cm
\epsffile{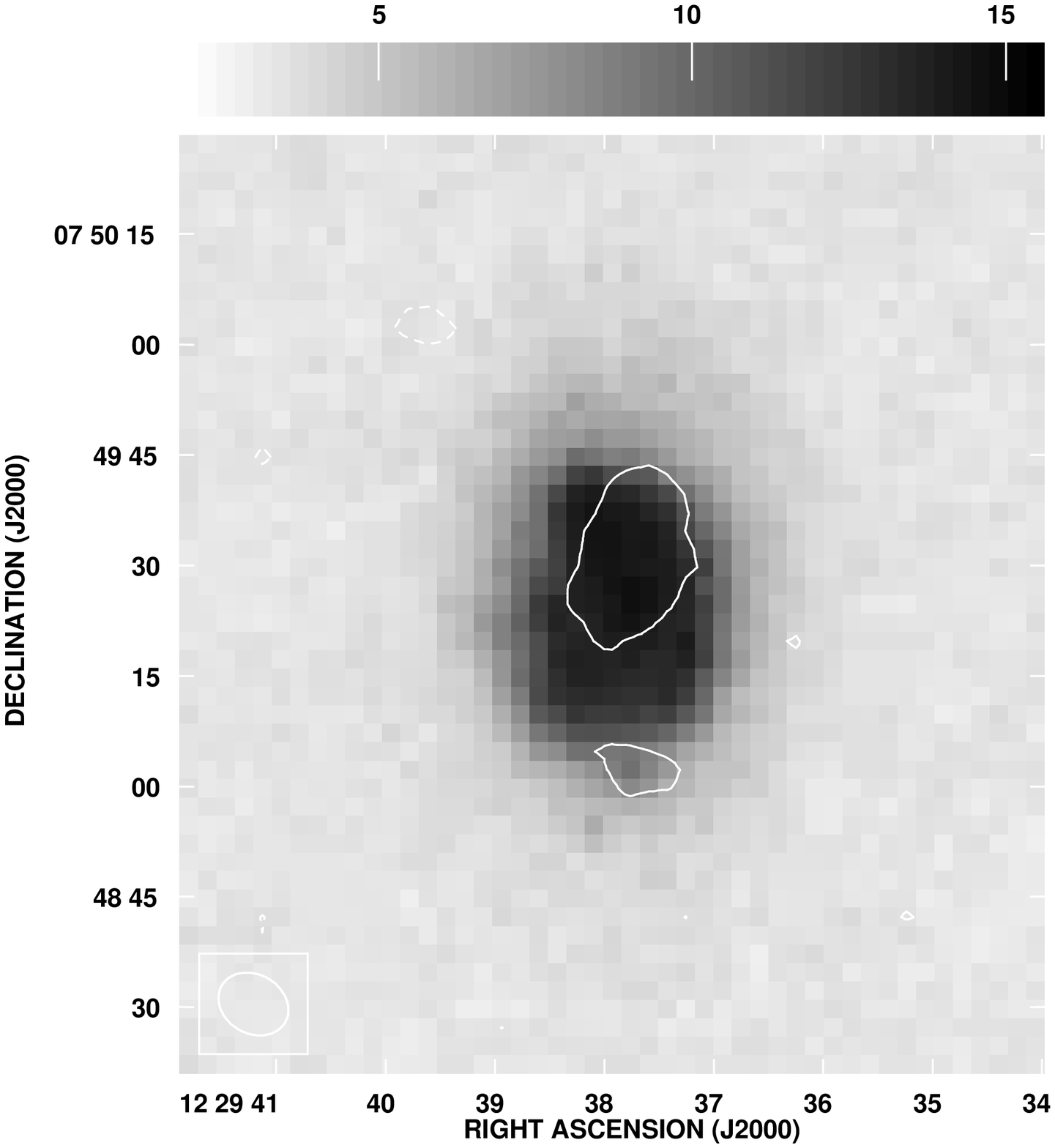}
(a)
\epsfxsize=7.3cm
\raisebox{0.4cm}{
\epsffile{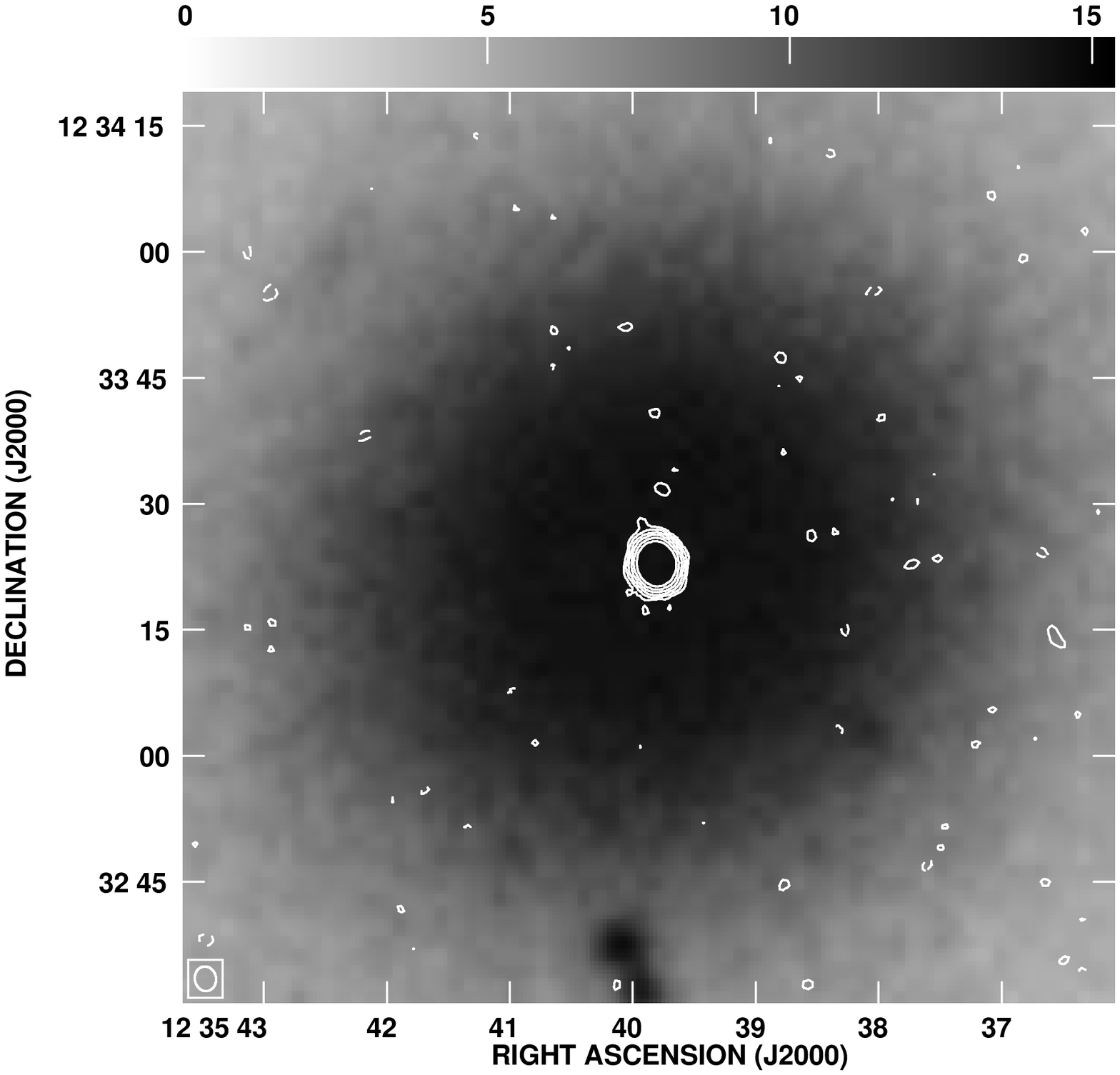}
(b)
}
}

\end{figure}

\begin{figure}
\figurenum{4}
\leavevmode
\centerline{
\epsfxsize=7.3cm
\raisebox{1.8cm}{
\epsffile{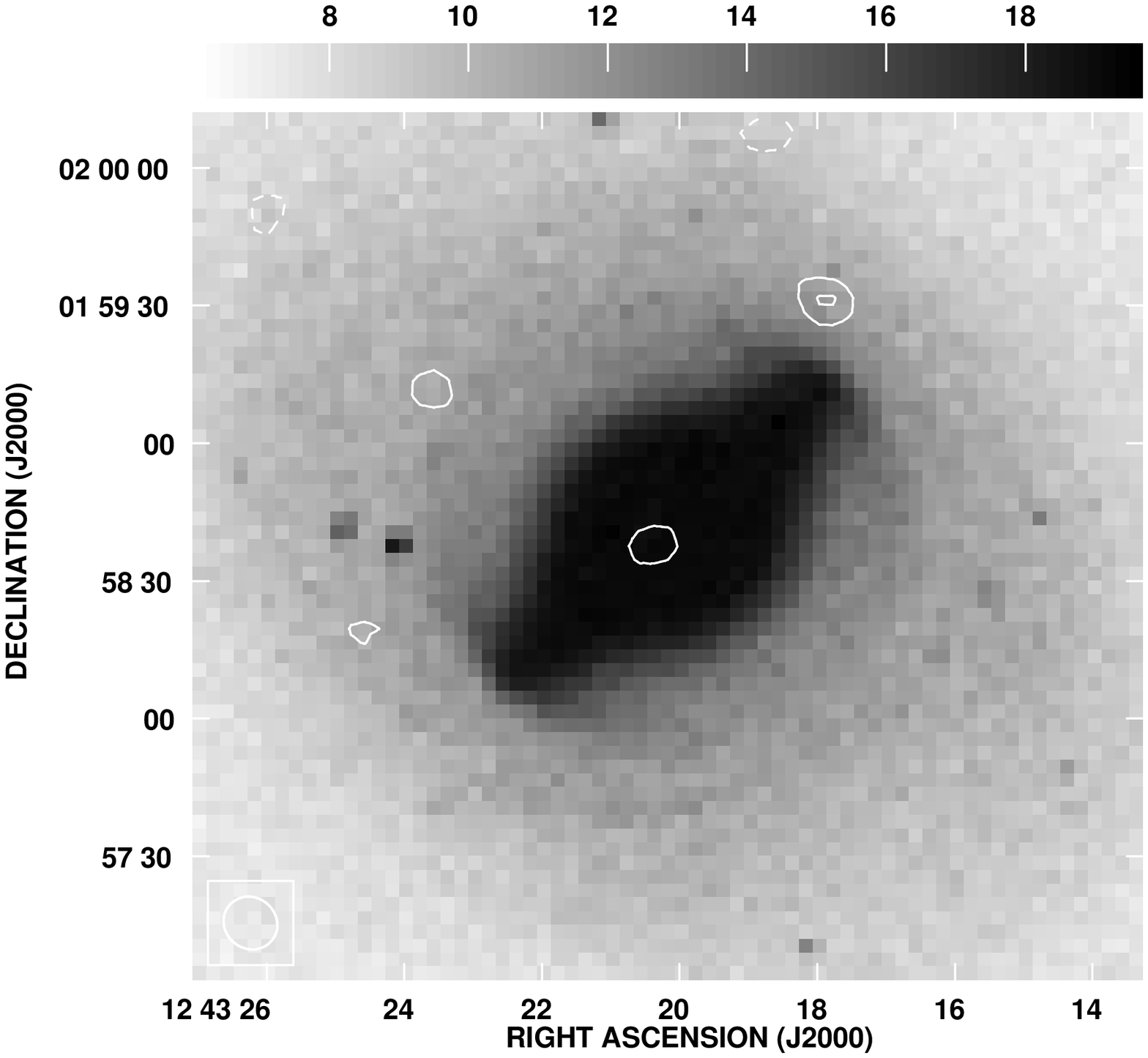}
(c)
}
\epsfxsize=7.3cm
\epsffile{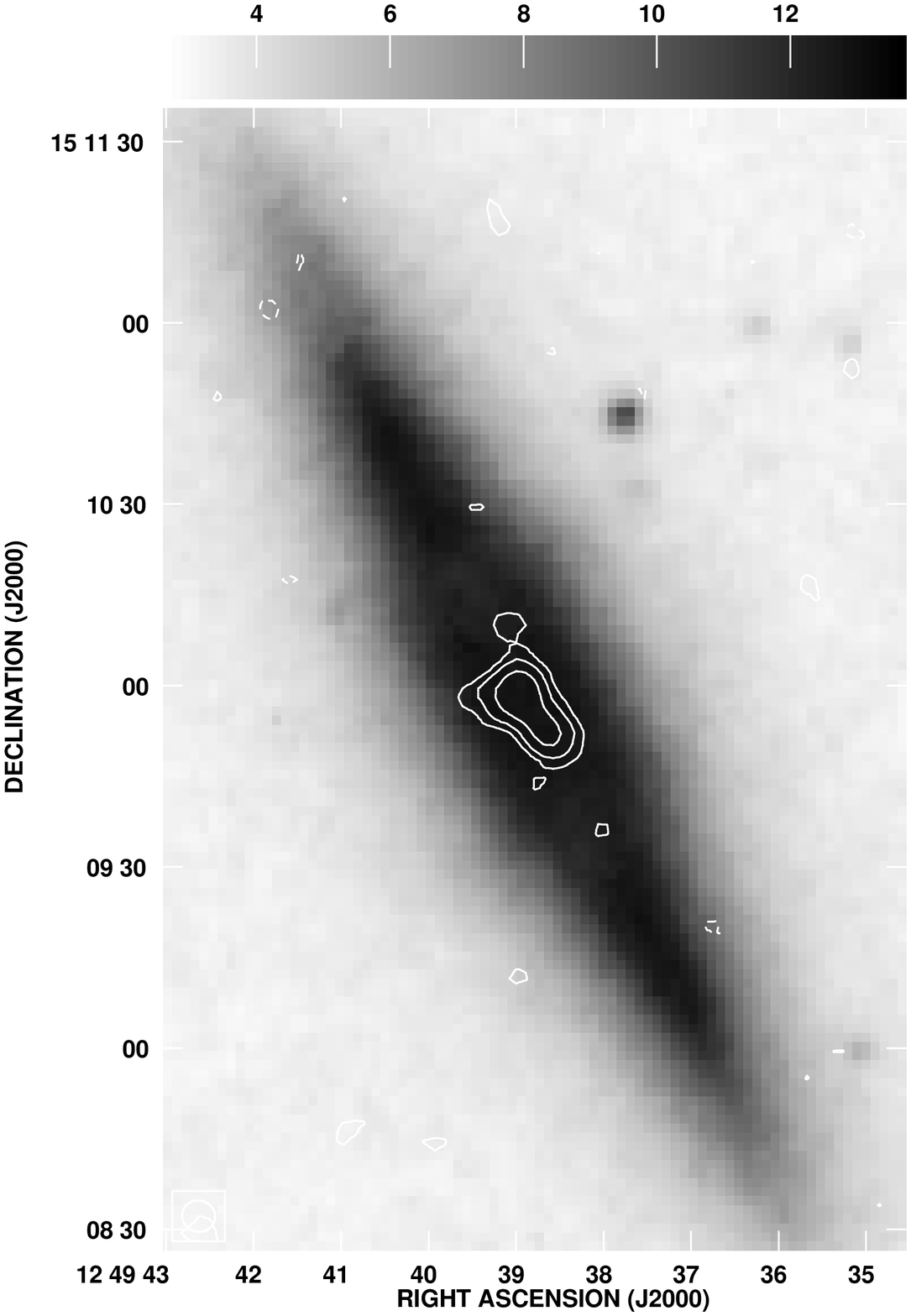}
(d)
}

\caption{As in Figure~1. {\bf (a)} NGC\,4470 (9\arcsecpoint95 $\times$ 7\arcsecpoint99), {\bf (b)} NGC\,4552
(2\arcsecpoint93 $\times$ 2\arcsecpoint54), {\bf (c)} NGC\,4643 (12\arcsecpoint04 $\times$ 11\arcsecpoint13) and {\bf (d)}
NGC\,4710 (5\arcsecpoint50 $\times$ 5\arcsecpoint16).  NGC\,4470 and NGC\,4710 are H\,II nuclei.}

\end{figure}

\clearpage

\begin{figure}
\leavevmode
\centerline{
\epsfxsize=7.3cm
\raisebox{1cm}{
\epsffile{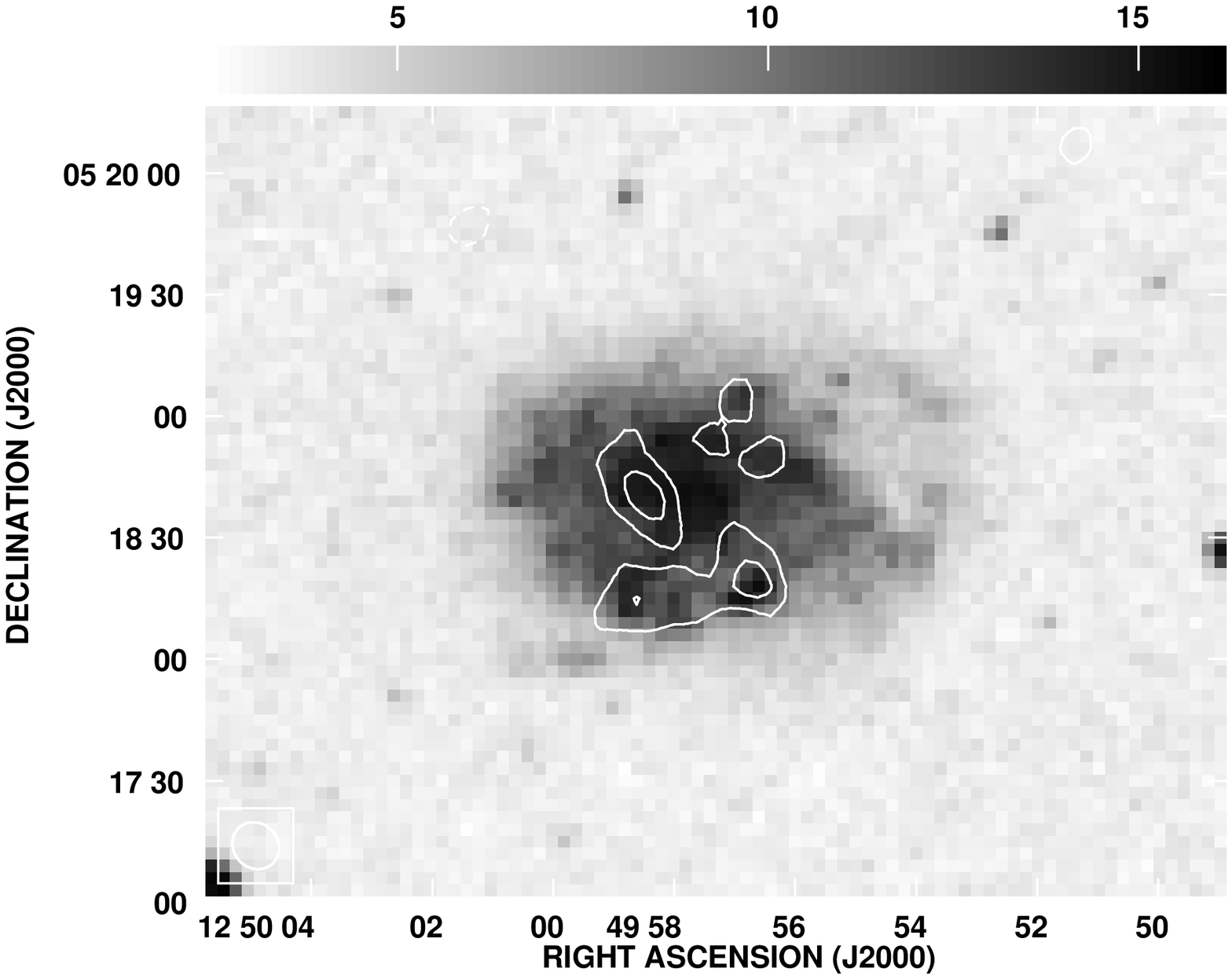}
(a)
}
\epsfxsize=7.3cm
\epsffile{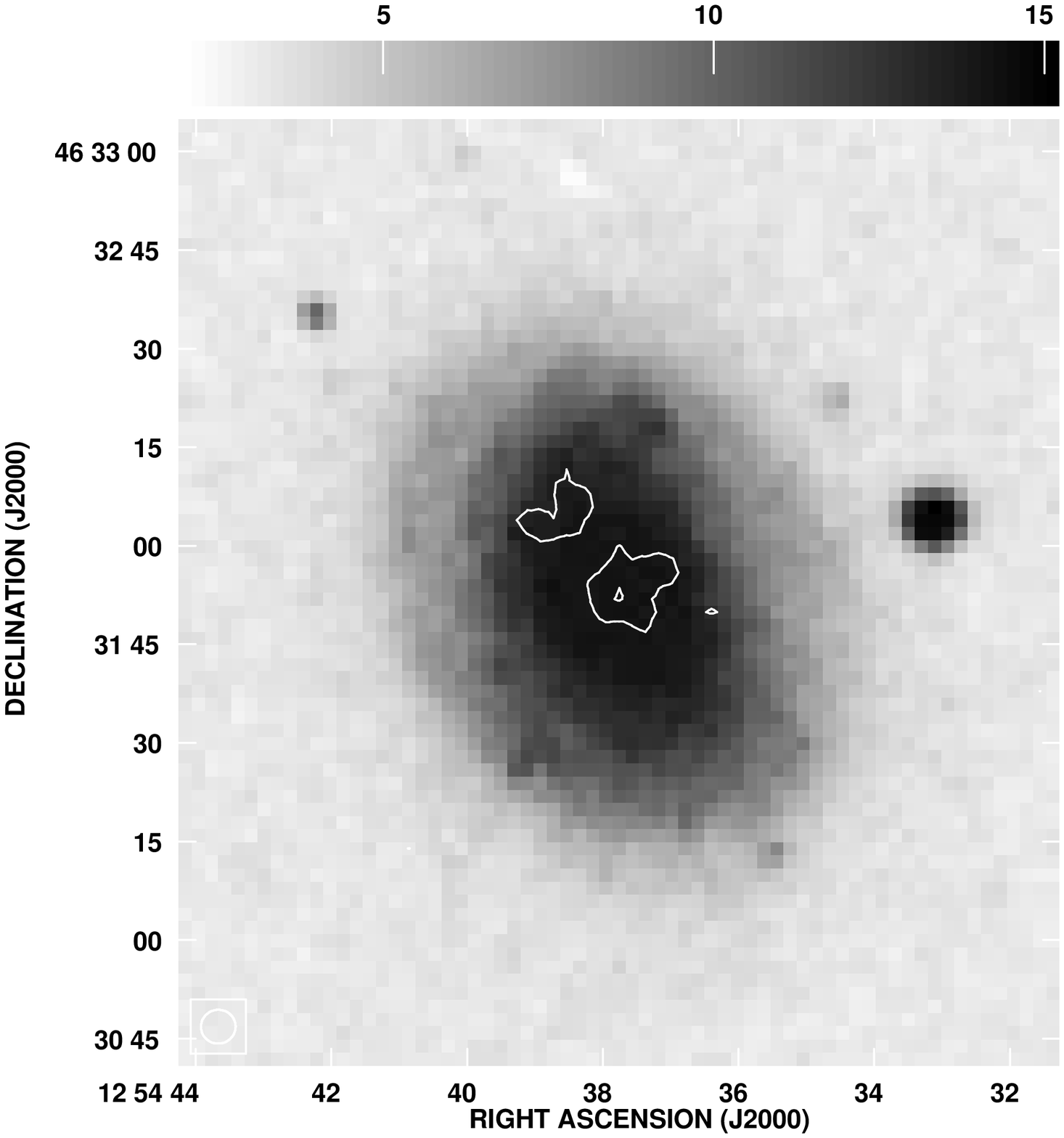}
(b)
}

\end{figure}

\begin{figure}
\figurenum{5}
\leavevmode
\centerline{
\epsfxsize=7.3cm
\epsffile{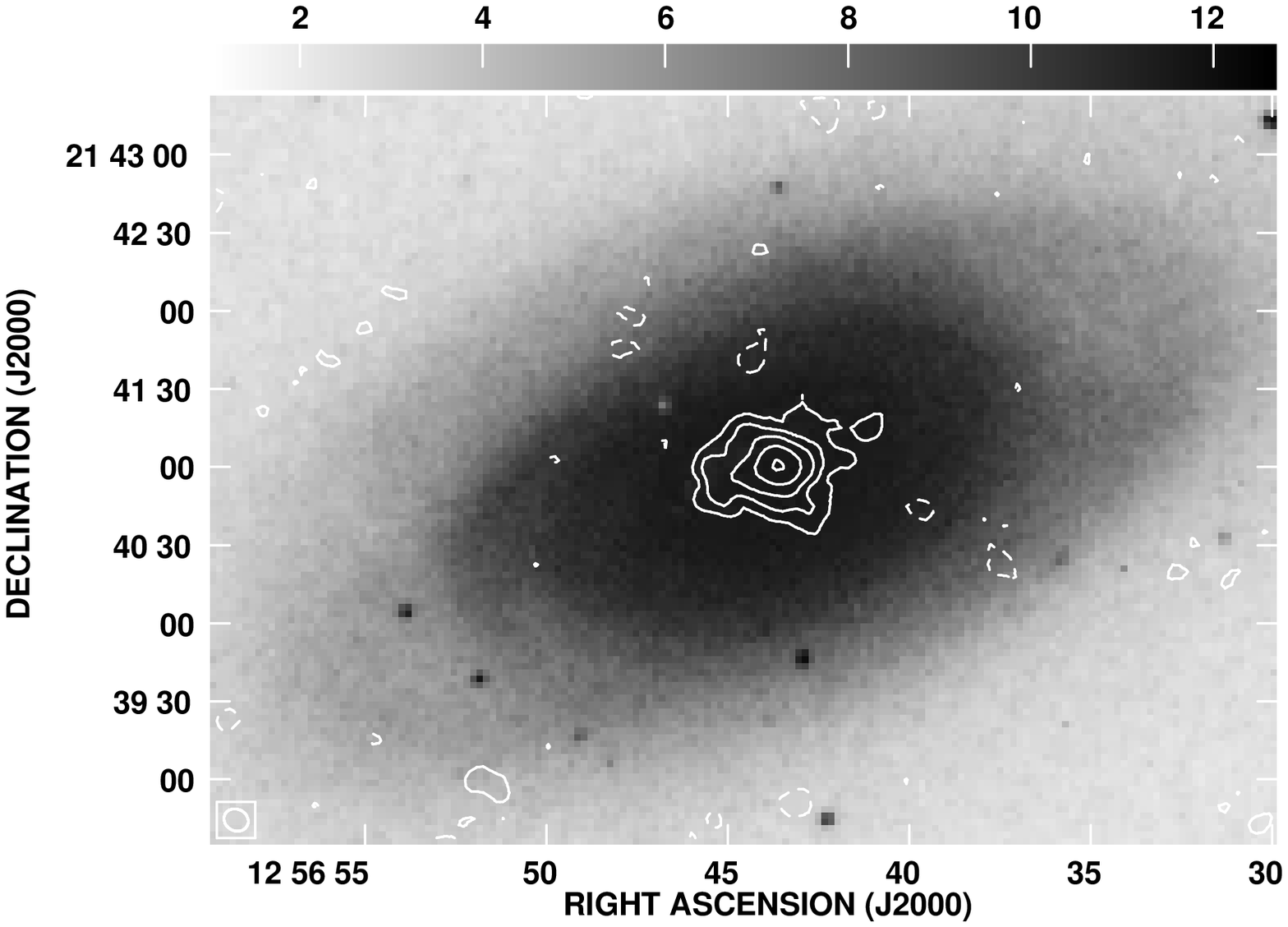}
(c)
\epsfxsize=7.3cm
\raisebox{0.6cm}{
\epsffile{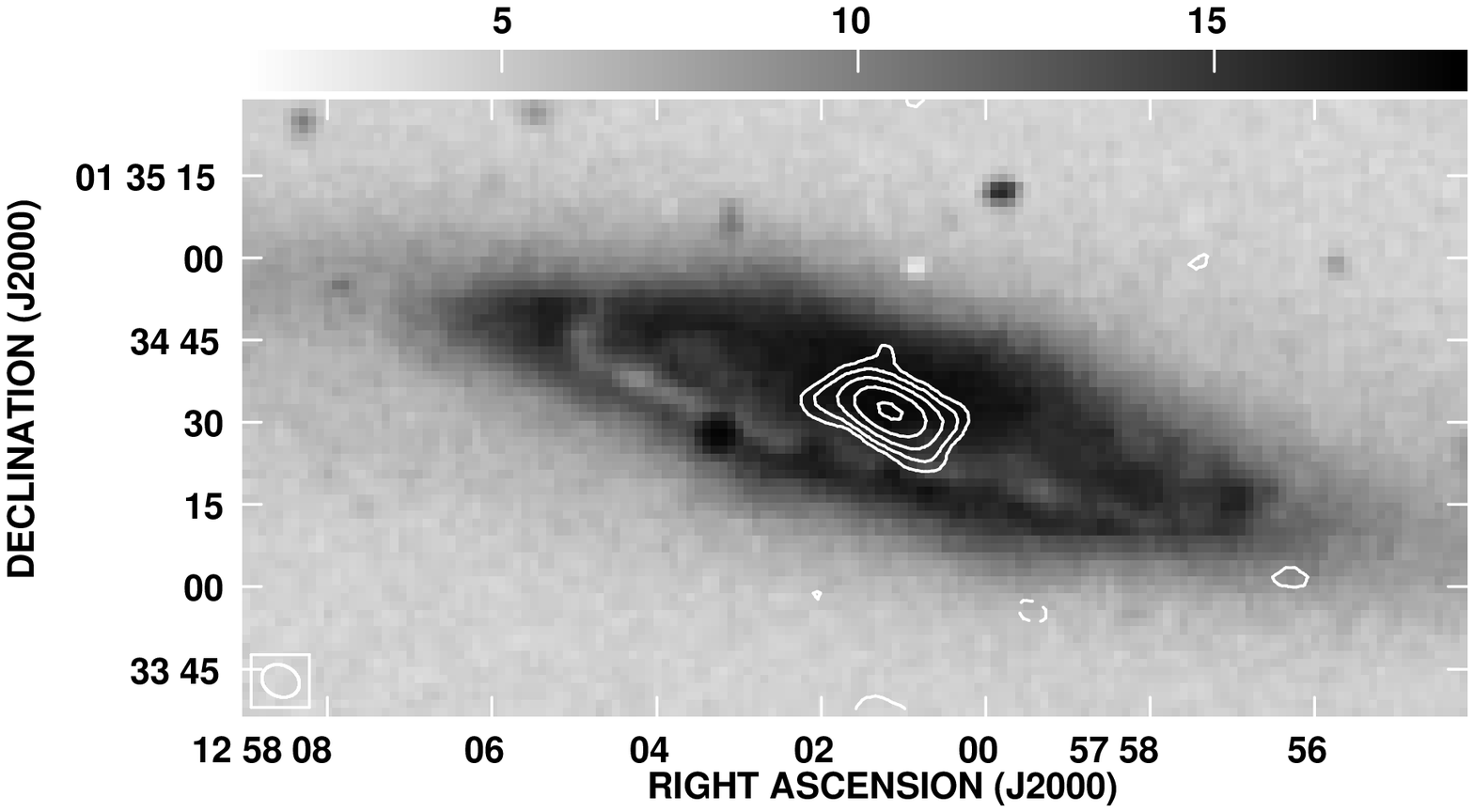}
(d)
}
}

\caption{As in Figure~1. {\bf (a)} NGC\,4713 (12\arcsecpoint12 $\times$ 10\arcsecpoint95), {\bf (b)} NGC\,4800
(5\arcsecpoint28 $\times$ 5\arcsecpoint15), {\bf (c)} NGC\,4826 (9\arcsecpoint62 $\times$ 8\arcsecpoint26) and {\bf (d)} NGC\,4845
(7\arcsecpoint08 $\times$ 5\arcsecpoint69).  NGC\,4845 is an H\,II nucleus.}

\end{figure}

\clearpage

\begin{figure}
\leavevmode
\centerline{
\epsfxsize=7.3cm
\epsffile{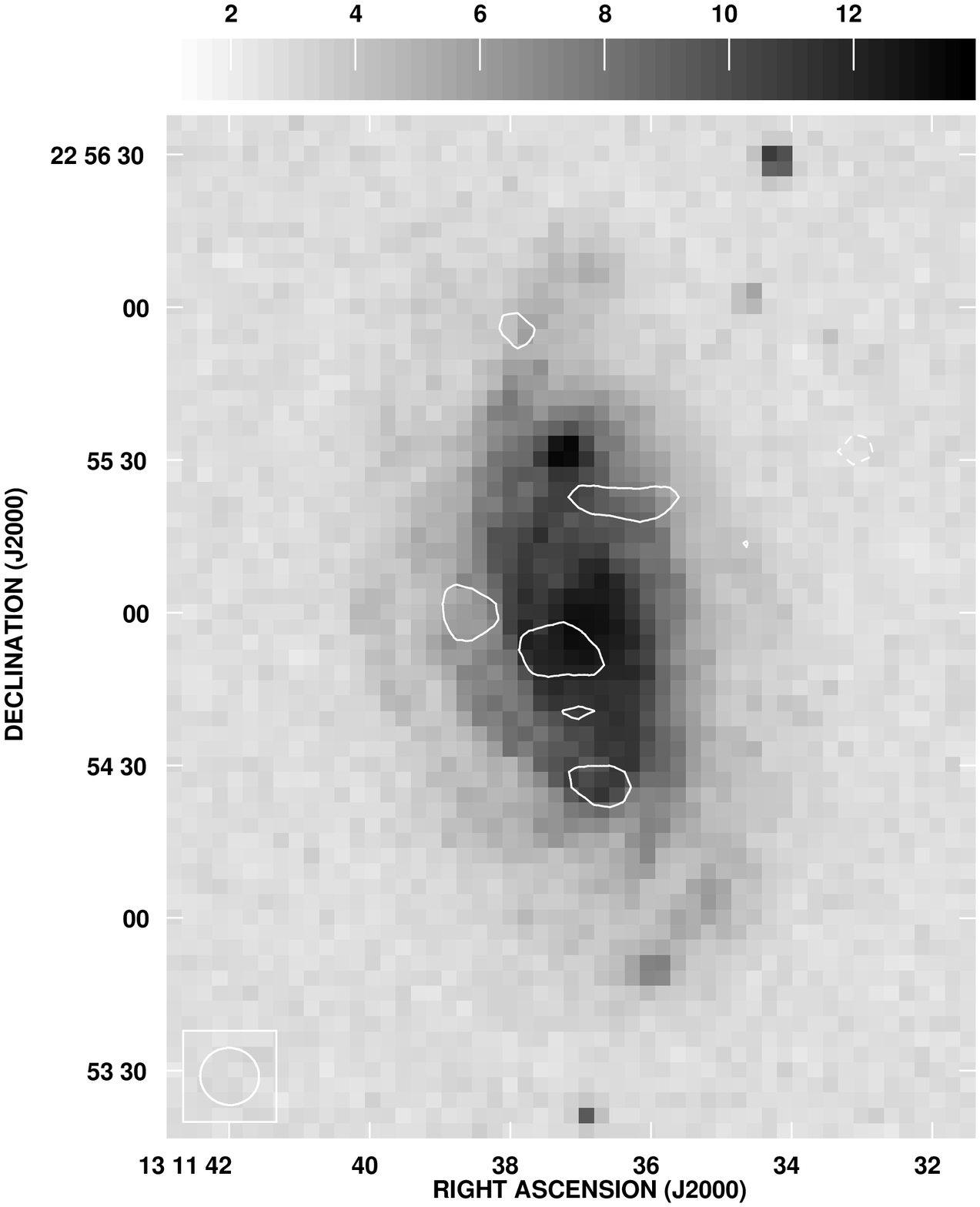}
(a)
\epsfxsize=7.3cm
\raisebox{1.1cm}{
\epsffile{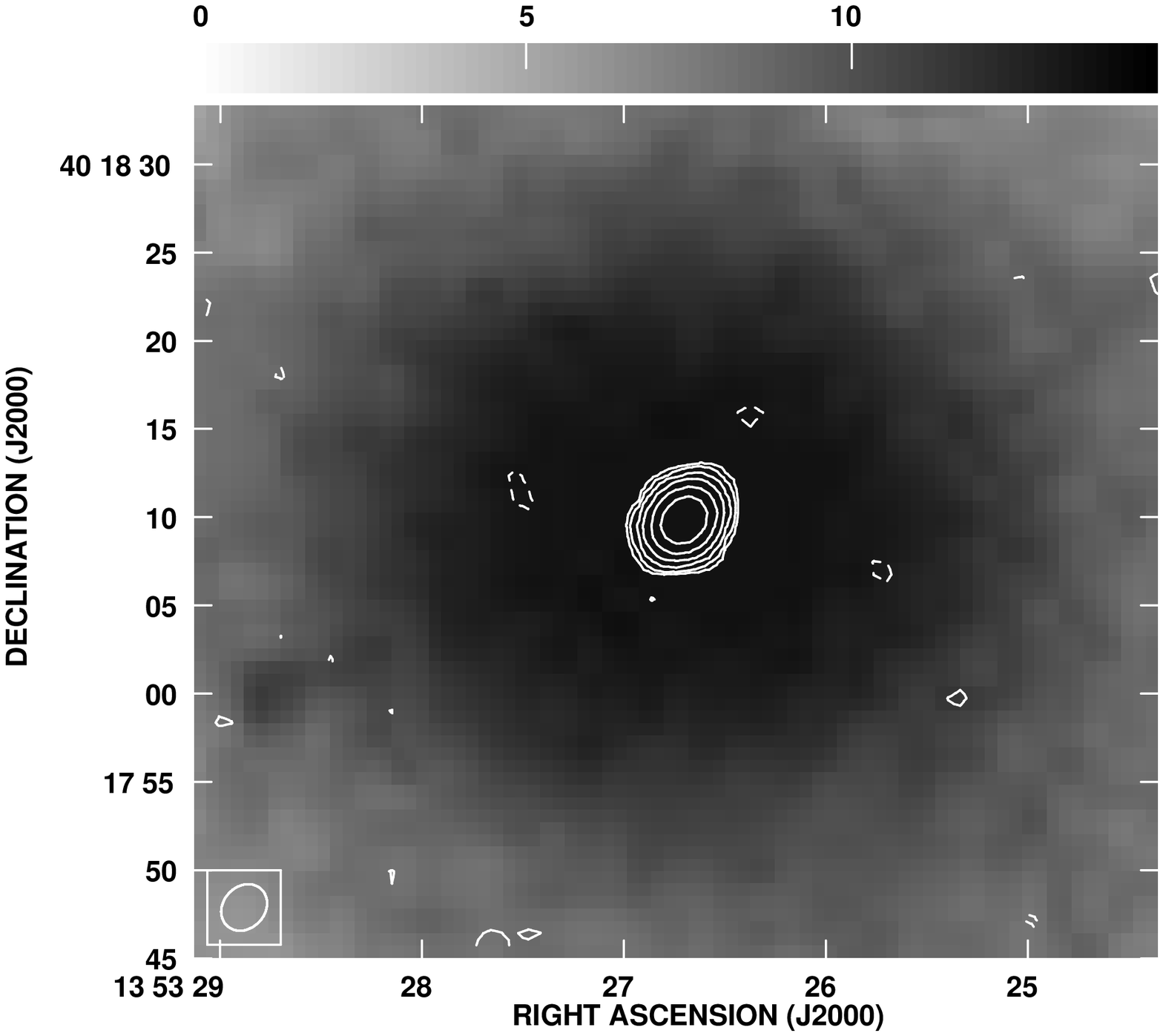}
(b)
}
}

\end{figure}

\begin{figure}
\figurenum{6}
\leavevmode
\centerline{
\epsfxsize=7.3cm
\raisebox{0.6cm}{
\epsffile{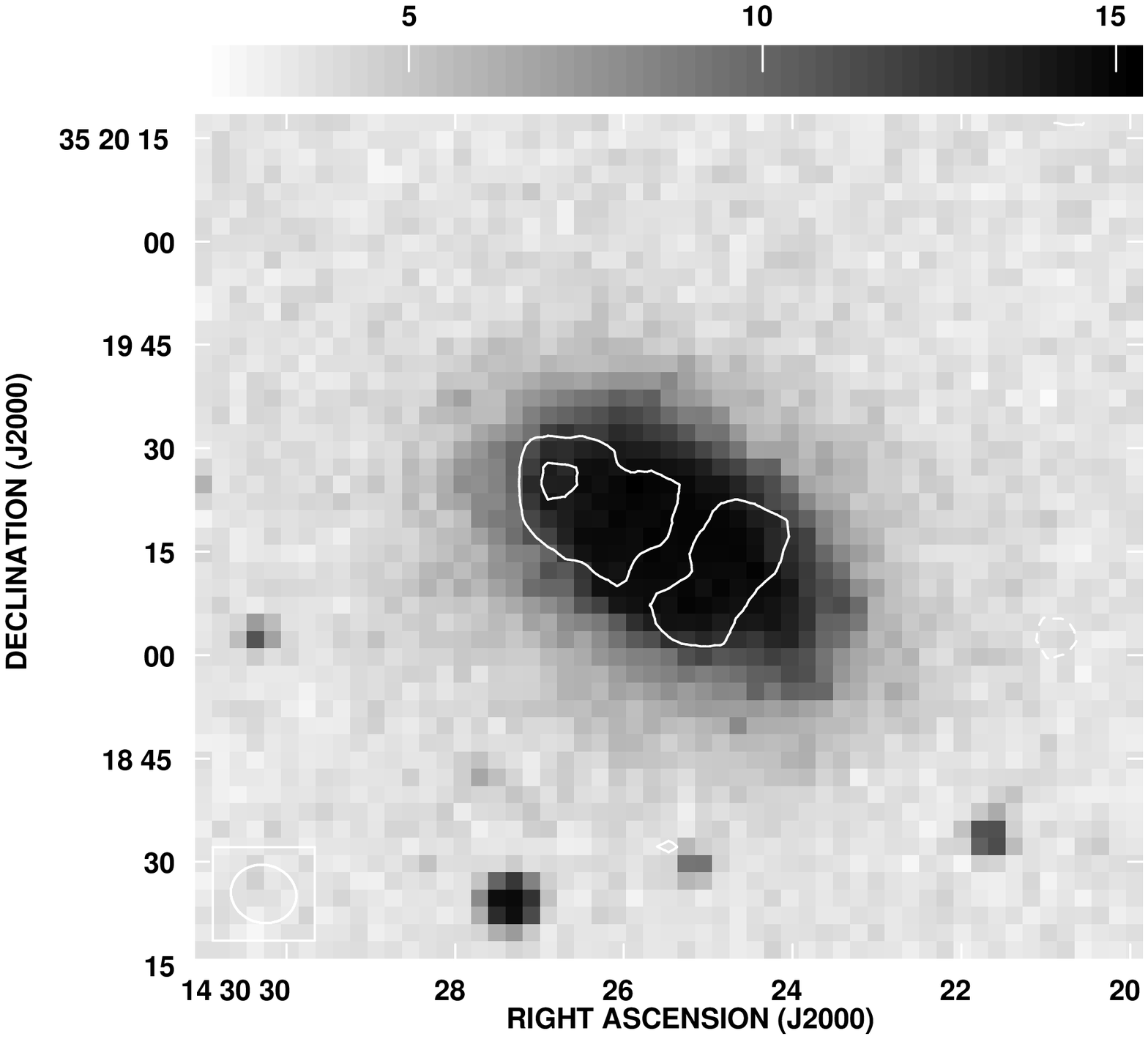}
(c)
}
\epsfxsize=7.3cm
\epsffile{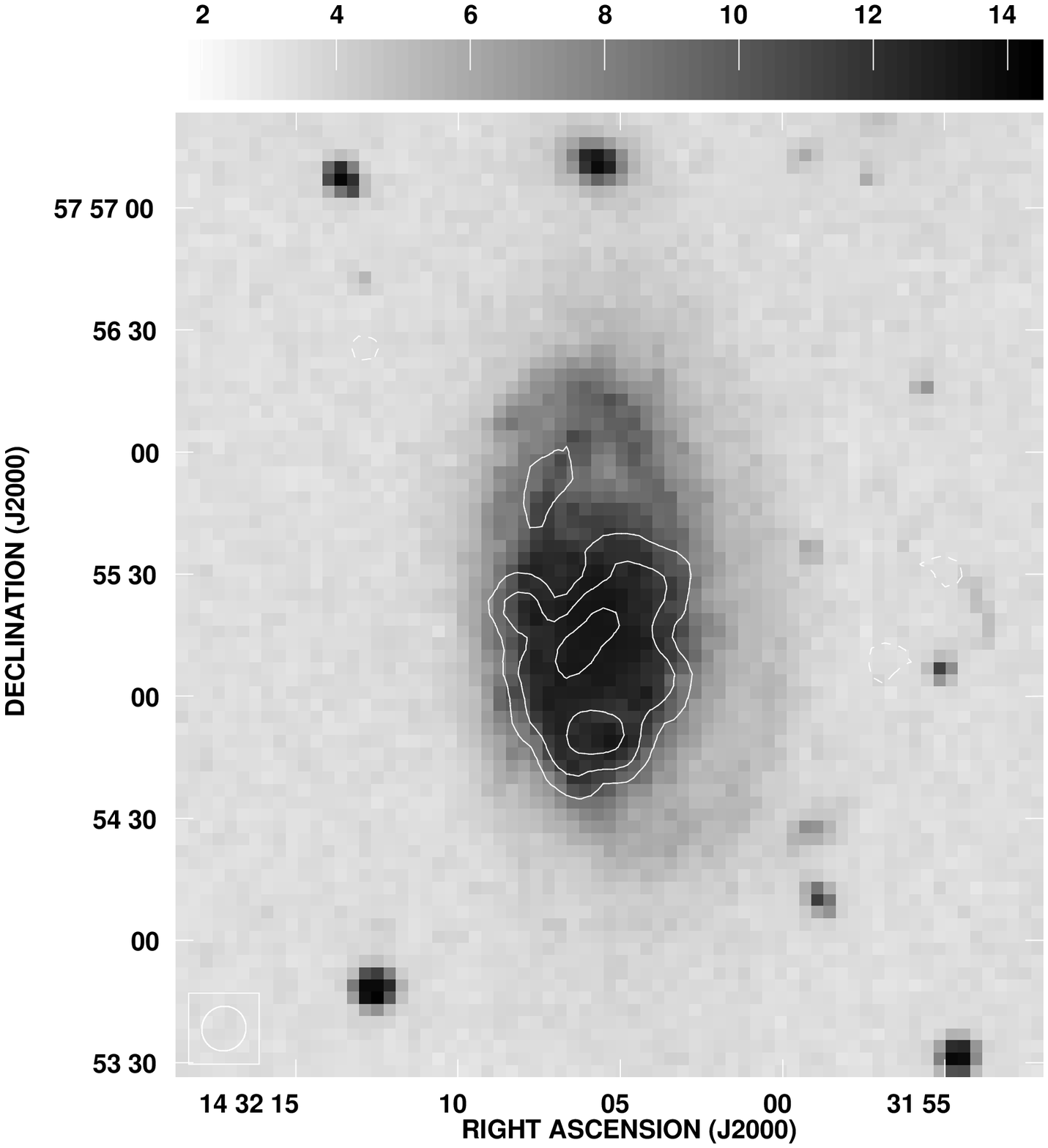}
(d)
}

\caption{As in Figure~1. {\bf (a)} NGC\,5012 (11\arcsecpoint59 $\times$ 11\arcsecpoint18), {\bf (b)} NGC\,5354
(2\arcsecpoint84 $\times$ 2\arcsecpoint37), {\bf (c)} NGC\,5656 (9\arcsecpoint54 $\times$ 8\arcsecpoint39) and {\bf (d)} NGC\,5678
(11\arcsecpoint09 $\times$ 10\arcsecpoint75).}

\end{figure}

\clearpage

\begin{figure}
\leavevmode
\centerline{
\epsfxsize=7.3cm
\epsffile{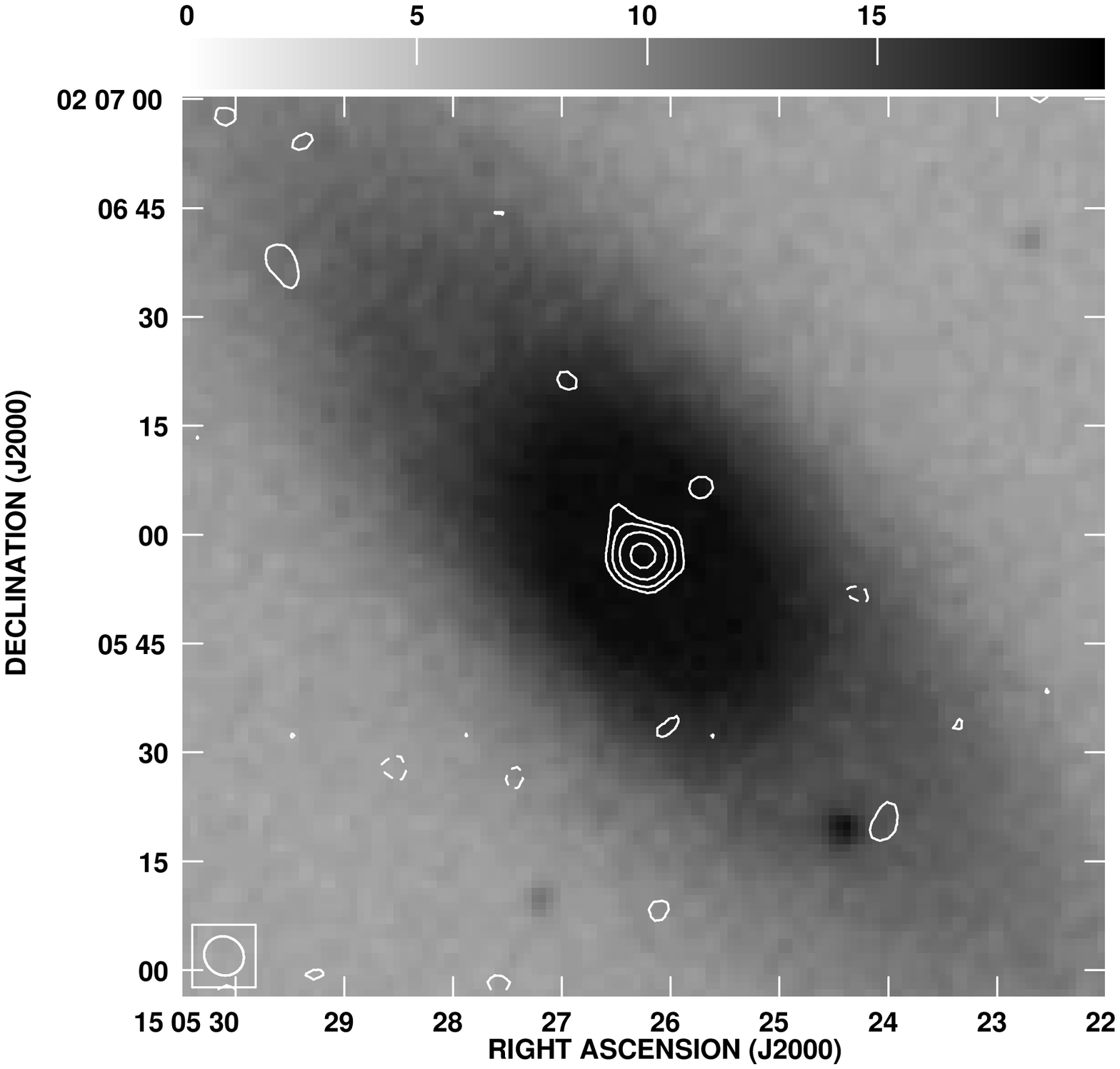}
(a)
\epsfxsize=7.3cm
\raisebox{0.3cm}{
\epsffile{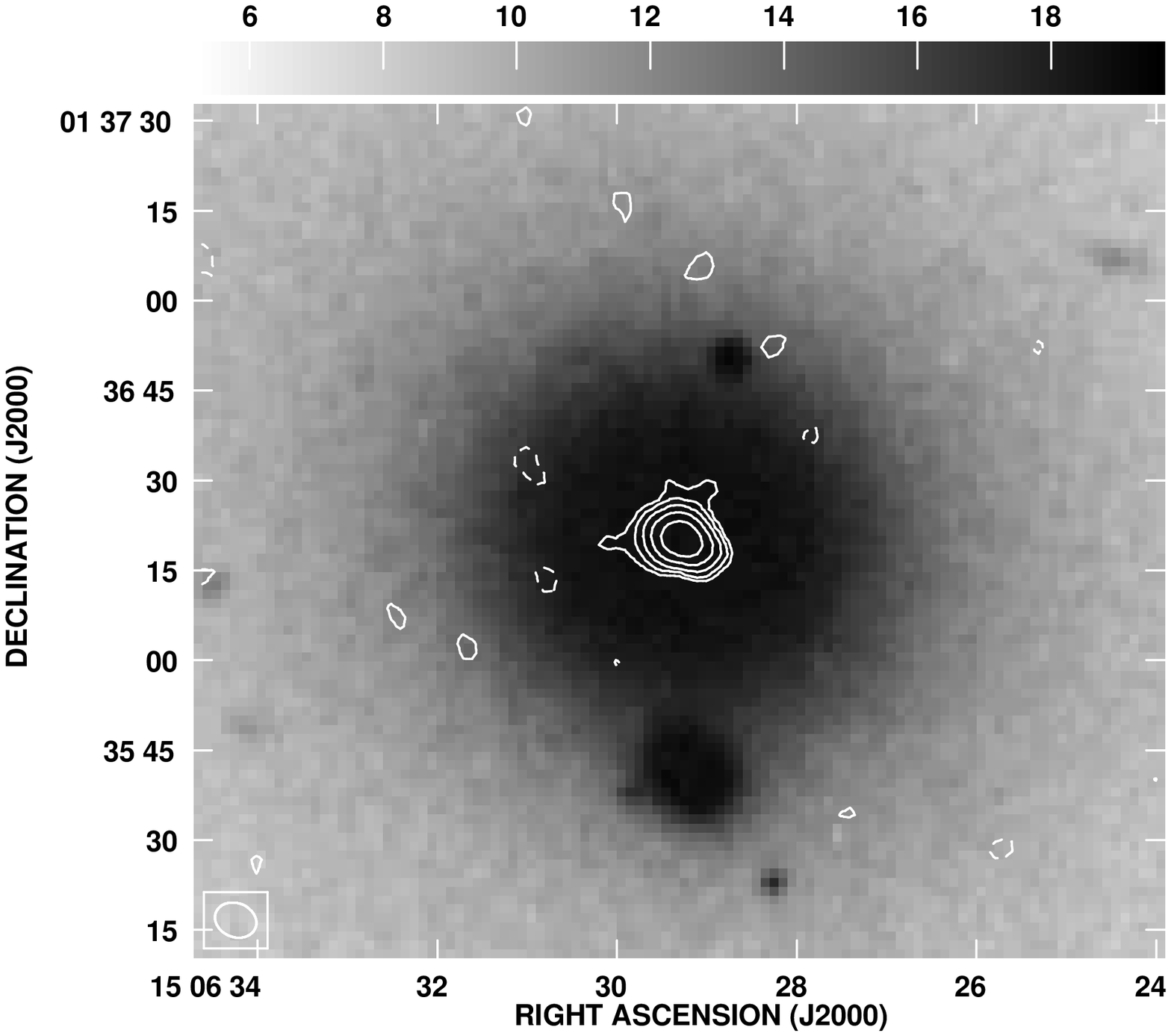}
(b)
}
}
\end{figure}

\begin{figure}
\figurenum{7}
\centerline{
\epsfxsize=7.3cm
\epsffile{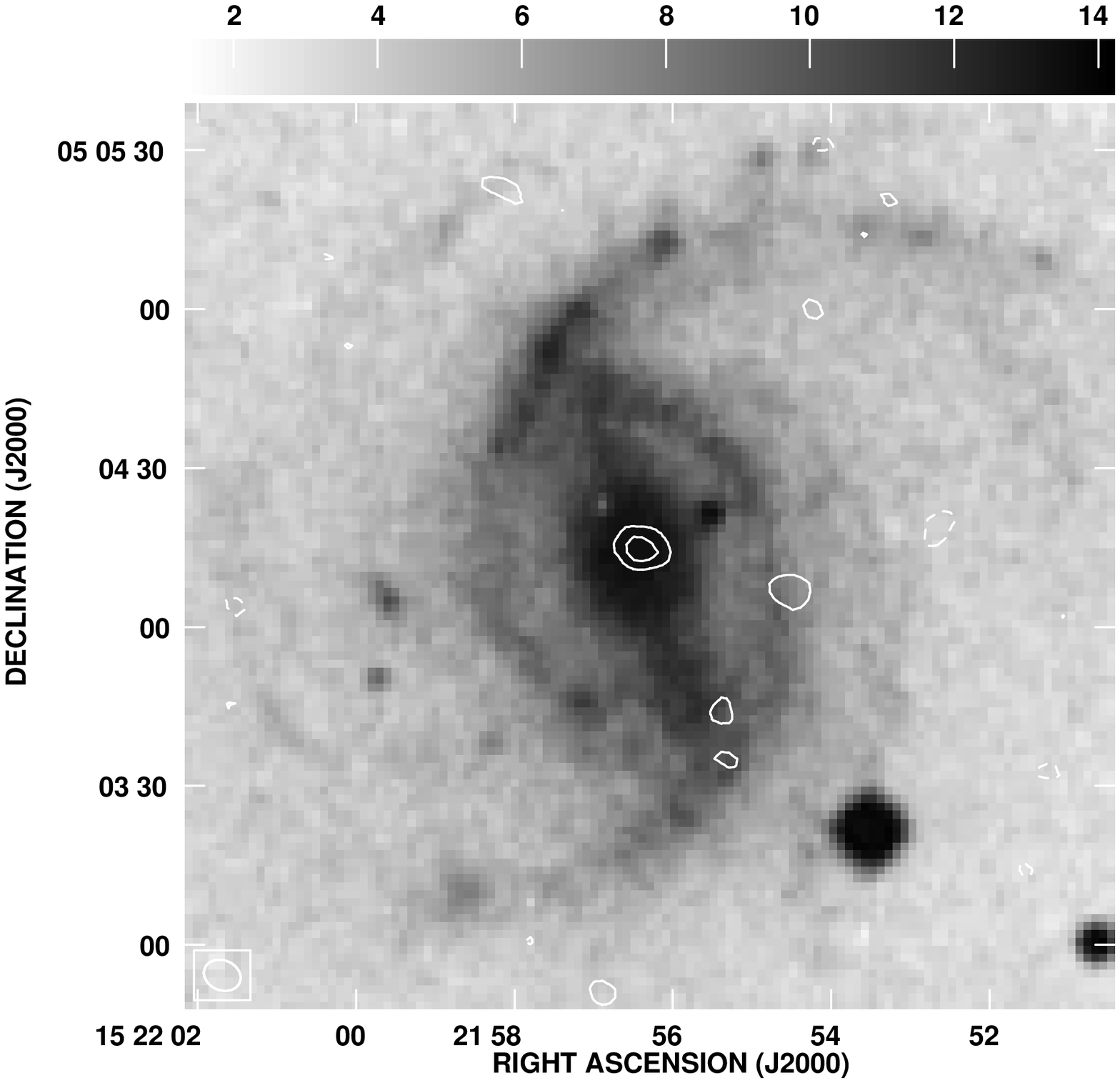}
(c)
}

\caption{As in Figure~1. {\bf (a)} NGC\,5838 (5\arcsecpoint62 $\times$ 5\arcsecpoint23), {\bf (b)} NGC\,5846
(7\arcsecpoint06 $\times$ 5\arcsecpoint67) and {\bf (c)} NGC\,5921 (7\arcsecpoint12 $\times$ 5\arcsecpoint63).}

\end{figure}


\begin{references}

\reference{}
Becker, R.~H., White, R.~L., \& Edwards, A.~L. 1991, 
\apjs, 75, 1

\reference{}
Becker, R.~H., White, R.~L., \& Helfand, D.~J. 1995,
\apj, 450, 559

\reference{}
Carral, P., Turner, J.~L., \& Ho, P.~T.~P. 1990, \apj, 362, 434

\reference{}
Clark, B.~G., 1980, \aap, 89, 377


\reference{}
Collison, P.~M., Saikia, D.~J., Pedlar, A., Axon, D.~J., \& Unger, S.~W. 1994,
\mnras, 268, 203
 
\reference{}
Condon, J.~J. 1983, \apjs, 53, 459

\reference{}
Condon, J.~J. 1987, \apjs, 63, 485

\reference{}
Condon, J.~J. 1992, ARA\&A, 30, 575

\reference{}
Condon, J.~J., \& Broderick, J.~J. 1988, \aj, 96, 30

\reference{}
Condon, J.~J., Condon, M.~A., Gisler, G., \& Puschell, J.~J. 1982, 
\apj, 252, 102

\reference{}
Condon, J.~J., Cotton, W.~D., Greisen, E.~W., Yin, Q.~F., 
Perley, R.~A., Taylor, G.~B., \& Broderick, J.~J. 1998a, \aj, 115, 1693

\reference{}
Condon, J.~J., Frayer, D.~T., \& Broderick, J.~J. 1991,
\aj, 101, 362

\reference{}
Condon, J.~J., Helou, G., Sanders, D.~B., \& Soifer, B.~T. 1990, 
\apjs, 73 359

\reference{}
Condon, J.~J., Yin, Q.~F., Thuan, T.~X., \& Boller, Th. 1998b,
\aj, 116, 2682

\reference{}
Cowan, J.~J., Romanishin, W., \& Branch, D. 1994, \apj, 436, L139

\reference{}
Dopita, M.~A., \& Sutherland, R.~S. 1995, \apj, 102, 161

\reference{}
Douglas, J.~N., Bash, F. N., Bozyan, F.~A., Torrence, G.~W., \& 
Wolfe, C. 1996, \aj, 111, 1945

\reference{}
Dressel, L.~L., \& Condon, J.~J. 1978, \apjs, 36, 53

\reference{}
Dumke, M.~K., Krause, M., Wielebinski, R., \& Klein, U. 1995,
\aap, 302, 691

\reference{}
Ekers, R.~D., \& Ekers, J.~A. 1973, \aap, 24, 247

\reference{}
Ekers, R.~D., Fanti, R., \& Miley, G.~K. 1983, \aap, 120, 297

\reference{}
Fabbiano, G., Gioia, I.~M., \& Trinchieri, G. 1989, \apj, 347, 127

\reference{}
Falcke, H., Wilson, A.~S., \& Ho, L.~C. 1997, in Relativistic
Jets in AGN, ed. M. Sikora \& M. Ostrowski, 13

\reference{}
Fanaroff, B.~L., \& Riley, J.~M. 1974, \mnras, 167, 31P

\reference{}
Filippenko, A.~V 1996, in The Physics of LINERs in View of Recent Observations, 
ed. M. Eracleous et al. (San Francisco: ASP), 17

\reference{}
Filippenko, A.~V., \& Terlevich, R. 1992, \apj, 397, L79

\reference{}
Fosbury, R.~A.~E., Melbold, U., Goss, W.~M., \& Dopita, M.~A. 1978, MNRAS, 183,
549

\reference{}
Gioia, I.~M., \& Fabbiano, G. 1987, \apjs, 63, 771

\reference{}
Giuricin, G., Fadda, D., \& Mezzetti, M. 1996, \apj, 468, 475

\reference{}
Gregory, P.~C., \& Condon, J.~J. 1991,
\apjs, 75, 1011

\reference{}
Haynes, M., \& Sramek, R.~A. 1975, \aj, 80, 673

\reference{}
Heckman, T.~M. 1980, \aap, 87, 152

\reference{}
Heckman, T.~M., Balick, B., \& Crane, P.~C. 1980, \aaps, 40, 295

\reference{}
Hewett, P.~C., Foltz, C.~B., \& Chaffee, F.~H. 1995, \aj, 109, 1498

\reference{}
Ho, L.~C. 1996, in The Physics of LINERs in View of Recent Observations, 
ed. M. Eracleous et al. (San Francisco: ASP), 103

\reference{}
Ho, L.~C. 1999a, \apj, 510, 631 

\reference{}
Ho, L.~C. 1999b, Advances in Space Research, 23 (5-6), 813

\reference{}
Ho, L.~C., Filippenko, A.~V., \& Sargent, W.~L.~W. 1993, \apj, 417, 63

\reference{}
Ho, L.~C., Filippenko, A.~V., \& Sargent, W.~L.~W. 1995, \apjs, 98, 477

\reference{}
Ho, L.~C., Filippenko, A.~V., \& Sargent, W.~L.~W. 1997a, \apjs, 112, 315

\reference{}
Ho, L.~C., Filippenko, A.~V., \& Sargent, W.~L.~W. 1997b, \apj, 487, 568

\reference{}
Ho, L.~C., Van Dyk, S.~D., Pooley, G.~G., Sramek, R.~A., \& Weiler, K.~W. 1999,
\aj, 118, 843

\reference{}
H\"ogbom, J.~A., 1974, \aaps, 15, 417

\reference{}
Hook, I.~M., McMahon, R.~G., Patnaik, A.~R., Browne, I.~W.~A., Wilkinson, 
P.~N., Irwin, M.~J.,\&  Hazard, C. 1995, \mnras, 273, L63

\reference{}
Hughes, P.~A., Aller, H.~D., \& Aller, M.~F. 1992, \apj, 396, 469

\reference{}
Hummel, E. 1980, \aaps, 41, 151

\reference{}
Hummel, E., Beck, R., \& Dettmar, R.-J. 1991,
\aaps, 87, 309

\reference{}
Hummel, E., \& Kotanyi, C.~G. 1982, \aap, 106, 183
 
\reference{}
Hummel, E., van der Hulst, J.~M., Keel, W.~C., \& Kennicutt, R.~C., Jr.
1987, \aaps, 70, 517

\reference{}
Jenkins, C.~R. 1982, \mnras, 200, 705

\reference{}
Jones, D.~L., Terzian, Y., \& Sramek, R.~A. 1981a, \apj, 246, 28

\reference{}
Jones, D.~L., Terzian, Y., \& Sramek, R.~A. 1981b, \apj, 247, L57

\reference{}
Kenney, J.~D.~P., Koopmann, R.~A., Rubin, V.~C., \& Young, J.~S. 1996,
\aj, 111, 152

\reference{}
Klein, U., \& Emerson, D.~T. 1981, \aap, 94, 29

\reference{}
M\"ollenhoff, C., Hummel, E., \& Bender, R. 1992,
\aap, 255, 35

\reference{}
Niklas, S., Klein, U., Braine, J., \& Wielebinski, R. 1995,
\aaps, 114, 21

\reference{}
Norris, R.~P., Allen, D.~A., Sramek, R.~A., Kesteven, M.~J., \& Troup, E.~R. 
1990, 
\apj, 359, 291

\reference{}
Patnaik, A.~R., Browne, I.~W.~A., Wilkinson, P.~N., \& Wrobel, J.~M. 1992, 
\mnras, 254, 655

\reference{}
Perley, R.~A., Schwab, F.~R., \& Bridle, A.~H. 1989, in Synthesis Imaging in 
Radio Astronomy: A Collection of Lectures from the Third NRAO Synthesis
Imaging Summer School (San Francisco: ASP), 6

\reference{}
Rengelink, R.~B., Tang, Y., de Bruyn, A.~G., Miley, G.~K., Bremer, M.~N., 
R\"ottgering, H.~J.~A., \& Bremer, M.~A.~R. 1997, \aaps, 124, 259

\reference{}
Reuter, H.-P., Krause, M., Wielebinski, R., \& Lesch, H. 1991,
\aap, 248, 12 


\reference{}
Sadler, E.~M., Jenkins, C.~R., \& Kotanyi, C.~G. 1989, \mnras, 240, 591

\reference{}
Saikia, D.~J., Pedlar, A., Unger, S.~W., \& Axon, D.~J. 1994,
\mnras, 270, 46

\reference{}
Schlickeiser, R., Werner, W., \& Wielebinski, R. 1984, \aap, 140, 227

\reference{}
Shields, J. C. 1992, \apj, 399, L27

\reference{}
Slee, O.~B., Sadler, E.~M., Reynolds, J.~E., \& Ekers, R.~D. 1994,
\mnras, 269, 928

\reference{}
Sramek, R.~A. 1975a, \aj, 80, 771

 \reference{}
Sramek, R.~A. 1975b, \apj, 198, L13

\reference{}
Terlevich, R., \& Melnick, J. 1985, \mnras, 213, 841

\reference{}
Turner, K.~C., Helou, G., \& Terzian, Y. 1988, PASP, 100, 452

\reference{}
Turner, J.~L., \& Ho, P.~T.~P. 1994,
\apj, 421, 122

\reference{}
Urbanik, M., Gr\"ave, R., \& Klein, U. 1985, \aap, 152, 291

\reference{}
Urbanik, M., Klein, U., \& Gr\"ave, R. 1986, \aap, 166, 107

\reference{}
van Breugel, W.~J.~M., Schilizzi, R.~T., Hummel, E., \& Kapahi, V.~K. 1981, 
\aap, 96, 310

\reference{}
van der Hulst, J.~M., Crane, \& P.~C., Keel, W.~C. 1981, \aj, 86, 1175

\reference{}
van der Kruit, P.~C. 1973a, \aap, 29, 231

\reference{}
van der Kruit, P.~C. 1973b, \aap,  29, 249

\reference{}
Van Dyk, S.~D., \& Ho, L.~C. 1998, in IAU Symp. 184, The Central Regions of
the Galaxy and Galaxies, ed. Y. Sofue (Dordrecht: Kluwer), 489

\reference{}
V\'eron-Cetty, M.-P., \& V\'eron, P. 1996, \aaps,  115, 97

\reference{}
Vila, M.~B., Pedlar, A., Davies, R.~D., Hummel, E., \& Axon, D.~J. 1990, 
\mnras, 242, 379

\reference{}
Weiler, K.~W., van der Hulst, J.~M., Sramek, R.~A., \& Panagia, N. 1981,
\apj, 243, L151

\reference{}
White, R.~L., \& Becker, R.~J. 1992,
\apjs, 79, 331

\reference{}
Wrobel, J.~M., \& Heeschen, D.~S. 1984, \apj, 287, 41

\reference{}
Wrobel, J.~M., \& Heeschen, D.~S. 1991,
\aj, 101, 148

\reference{}
Zirbel, E.~L., \& Baum, S.~A. 1995, \apj, 448, 521 
 
\end{references}
\end{document}